\documentclass[aps,prd,10pt,twocolumn,nofootinbib,superscriptaddress,preprintnumbers,floatfix]{revtex4-1}
\usepackage{amsfonts,amsmath,amssymb,amsthm,braket,nicefrac}
\usepackage{booktabs,array}
\usepackage[utf8]{inputenc}
\usepackage{float,graphicx,hhline}
\usepackage{algorithm,algpseudocode}
\usepackage{appendix}
\usepackage[dvipsnames]{xcolor}
\usepackage{mathtools}
\usepackage{multirow}
\usepackage{tikz-cd}
\usepackage{slashed}
\usepackage{soul}
\usepackage{pifont}
\usepackage[caption=false]{subfig}
\usepackage[export]{adjustbox}
\usepackage[normalem]{ulem}
\usepackage{hyperref}
\usepackage{cleveref}
\usepackage{listings}
\usepackage{comment}
\usepackage{pgf}

\AtBeginDocument{
\heavyrulewidth=.08em
\lightrulewidth=.05em
\cmidrulewidth=.03em
\belowrulesep=.65ex
\belowbottomsep=0pt
\aboverulesep=.4ex
\abovetopsep=0pt
\cmidrulesep=\doublerulesep
\cmidrulekern=.5em
\defaultaddspace=.5em
}

\usepackage{hyperref}%
\hypersetup{
    pdftitle={many pions},
    colorlinks=true,     %
    linkcolor=blue,      %
    citecolor=blue,      %
    filecolor=blue,      %
    urlcolor=blue        %
}

\definecolor{codegreen}{rgb}{0,0.6,0}
\definecolor{codegray}{rgb}{0.5,0.5,0.5}
\definecolor{codepurple}{rgb}{0.58,0,0.82}
\definecolor{backcolour}{rgb}{0.95,0.95,0.92}

\lstdefinestyle{mystyle}{
    backgroundcolor=\color{backcolour},   
    commentstyle=\color{codegreen},
    keywordstyle=\color{magenta},
    numberstyle=\tiny\color{codegray},
    stringstyle=\color{codepurple},
    basicstyle=\ttfamily\footnotesize,
    breakatwhitespace=false,         
    breaklines=true,                 
    captionpos=b,                    
    keepspaces=true,                 
    numbers=none,                    
    numbersep=5pt,                  
    showspaces=false,                
    showstringspaces=false,
    showtabs=false,                  
    tabsize=2
}

\DeclareMathAlphabet{\mathbbold}{U}{bbold}{m}{n}

\DeclareMathOperator*{\argmin}{arg\,min}

\begin{document}

\title{QCD constraints on isospin-dense matter and the nuclear equation of state}

\newcommand{\getMITAffiliation}{\affiliation{Center for Theoretical Physics, Massachusetts Institute of Technology, Cambridge, MA 02139, USA}}
\newcommand{\getIAIFIAffiliation}{\affiliation{The NSF AI Institute for Artificial Intelligence and Fundamental Interactions}}

\author{Ryan Abbott}
\getMITAffiliation
\getIAIFIAffiliation

\author{William Detmold}
\getMITAffiliation
\getIAIFIAffiliation

\author{Marc~Illa}
\affiliation{InQubator for Quantum Simulation (IQuS), Department of Physics, University of Washington, Seattle, WA 98195, USA}

\author{Assumpta~Parre\~no}
\affiliation{Departament de F\'{\i}sica Qu\`{a}ntica i Astrof\'{\i}sica and Institut de Ci\`{e}ncies del Cosmos,	Universitat de Barcelona, Mart\'{\i} i Franqu\`es 1, E08028, Spain}

\author{Robert J. Perry}
\affiliation{Departament de F\'{\i}sica Qu\`{a}ntica i Astrof\'{\i}sica and Institut de Ci\`{e}ncies del Cosmos,	Universitat de Barcelona, Mart\'{\i} i Franqu\`es 1, E08028, Spain}

\author{Fernando Romero-L\'opez}
\getMITAffiliation
\getIAIFIAffiliation

\author{Phiala~E.~Shanahan } 
\getMITAffiliation
\getIAIFIAffiliation

\author{Michael~L.~Wagman}
\affiliation{Fermi National Accelerator Laboratory, Batavia, IL 60510, USA}

\collaboration{NPLQCD collaboration}

\preprint{FERMILAB-PUB-24-0333-T}
\preprint{MIT-CTP/5730}

\begin{abstract}
Understanding the behavior of dense hadronic matter is a central goal in nuclear physics as it governs the nature and dynamics of astrophysical objects such as supernovae and neutron stars. Because of the non-perturbative nature of quantum chromodynamics (QCD), little is known rigorously about hadronic matter in these extreme conditions.
Here, lattice QCD calculations are used to compute thermodynamic quantities and the equation of state of QCD over a wide range of isospin chemical potentials with controlled systematic uncertainties.
Agreement is seen with chiral perturbation theory when the chemical potential is small.
Comparison to perturbative QCD at large chemical potential allows for an estimate of the gap in the superconducting phase, and this quantity is seen to agree with perturbative determinations. Since the partition function for an isospin chemical potential, $\mu_I$, bounds the partition function for a baryon chemical potential  $\mu_B=3\mu_I/2$, these calculations also provide rigorous non-perturbative QCD bounds on the symmetric nuclear matter equation of state over a wide range of baryon densities for the first time.
\end{abstract}
\maketitle

The determination of the internal structure of neutron stars presents a long-standing and important challenge for nuclear theory. Since neutron stars were first predicted in the 1930s and observed 30 years later, many models for the structure of their interiors have been proposed, including various phases of nuclear matter, mesonic condensates and hyperonic matter, and deconfined quark cores \cite{Ozel:2016oaf,Oertel:2016bki,Baym:2017whm,Mannarelli:2019hgn,Burgio:2021vgk,Lattimer:2021emm}. As observational data and terrestrial probes of the relevant nuclear densities are not sufficiently constraining, most of these possibilities for the neutron star equation of state (EoS) remain viable. From a theoretical perspective, it is expected that neutron star interiors can be described by the Standard Model of particle physics, however in a regime where the strong interactions are non-perturbative. The numerical technique of lattice quantum chromodynamics (LQCD) is applicable at such large couplings, but is beset by a notorious sign problem at nonzero baryon chemical potential \cite{deForcrand:2009zkb}, prohibiting its direct application. Consequently, theoretical approaches to the nuclear EoS are based on models and interpolations between phenomenological constraints from nuclear structure and perturbative QCD (pQCD) calculations that are valid at asymptotically large chemical potentials (see for example, Refs.~\cite{Annala:2023cwx,Semposki:2024vnp,Han:2022rug,Jiang:2022tps,Raaijmakers:2019dks,Huth:2020ozf,vanDalen:2005sk,Typel:2005ba,Kolomeitsev:2004ff,Liu:2001iz,Leonhardt:2019fua}). In light of this, any rigorous information that can impact such analyses is of paramount importance.

In this work, the first non-perturbative QCD constraint on the nuclear EoS with complete quantification of systematic uncertainties is presented. These calculations build upon the proof-of-principle, single lattice spacing results of Ref.~\cite{Abbott:2023coj} with improved methodology, increased statistical precision, an extrapolation to the continuum limit, and an interpolation to the physical quark masses, enabling a systematically-controlled result to be achieved for the first time. The pressure and other thermodynamic properties of low-temperature isospin-dense matter are determined over a wide range of densities and chemical potentials, spanning all scales from hadronic to perturbatively-coupled. At small values of the isospin chemical potential, $\mu_I$, the results agree with chiral perturbation theory ($\chi$PT) \cite{son_qcd_2001,Kogut:2001id} at next-to-leading order (NLO) \cite{Adhikari:2019zaj,Andersen:2023ofv}. At large $\mu_I$, the results are seen to agree with pQCD with pairing contributions \cite{Fujimoto:2023mvc}. The comparison of next-to-next-to-leading-order (NNLO) pQCD predictions \cite{Freedman:1976xs,Freedman:1976dm,Freedman:1976ub,Baluni:1977ms,Kurkela:2009gj,Fujimoto:2023mvc} for the pressure (partial next-to-next-to-next-to-leading-order results are also available \cite{Gorda:2021znl}) with the continuum limit of the LQCD calculations provides a determination of the superconducting gap as a function of $\mu_I$. This is seen to agree with the leading-order perturbative calculation of the pairing gap \cite{Fujimoto:2023mvc}, but is more precise.
The speed of sound in isospin-dense matter is also seen to significantly exceed the conformal limit of $c_s^2/c^2\leq1/3$ over a wide range of $\mu_I$. A Bayesian model mixing approach that combines $\chi$PT, LQCD and pQCD provides a determination of the zero-temperature EoS for isospin-dense QCD matter valid at all values of the isospin chemical potential.

These results provide constraints for phenomenological frameworks seeking to describe the QCD phase diagram \cite{MUSES:2023hyz}, and
from simple path-integral relations \cite{Cohen:2003ut,Cohen:2004qp,Fujimoto:2023unl,Moore:2023glb}, the determination of the pressure in isospin-dense matter provides a non-perturbative, model-independent bound on the pressure of isospin-symmetric QCD matter at nonzero baryon chemical potential and hence on the nuclear EoS. 
The current results therefore provide a systematically-controlled QCD bound at all densities, and the impact on nuclear phenomenology is briefly discussed.

\emph{Thermodynamic relations:}
Thermodynamic quantities are accessed in this work  by building an approximation
to the grand canonical partition function, valid
at low-temperature. The grand canonical partition function  is defined at a temperature, $T=1/\beta$, and isospin chemical potential, $\mu_I$, by
\begin{equation}
Z(\beta, \mu_I) = 
\sum_{s} e^{-\beta (E_s - \mu_I I_z(s))},
\label{eq:GCPF}
\end{equation}
where the sum is over all states, $s$,  and $E_s$ and $I_z(s)$ correspond to the energy and $z$-component of isospin of a given state, respectively. Since states of different $I_z$ but the same $I$ are approximately degenerate,  
contributions from states with $I_z< I$ are suppressed by ${\cal O}(e^{-\beta\mu_I})$ relative to those with $I_z=I$ and can therefore be neglected.
Additionally, only the ground state for each isospin contributes at low temperature. The summation can therefore be approximated in terms of these $I_z=I$ ground states, which can be labeled by their isospin charge $n=I=I_z$, and truncated at some $n_{\rm max}$ giving
\begin{equation}
    \label{eq:Zdef}
    Z(\beta\to \infty, \mu_I) \simeq \sum_{n=0}^{n_{\rm max}} e^{-\beta (E_n - \mu_I n)}.
\end{equation}
$E_0=0$ is chosen, and this approximation is valid for values of $\mu_I$ such that the truncation at $n_{\rm max}$ does not affect the result significantly.

For an observable $\mathcal{O}$ that only depends on the energy and isospin charge of the system, the thermodynamic expectation value of $\mathcal{O}$ can be computed as
\begin{equation}
    \label{eq:obsdef}
    \braket{\mathcal{O}(E,n)}_{\beta,\mu_I} = \frac{1}{Z(\beta, \mu_I)} \sum_n \mathcal{O}(E_n,n) e^{-\beta (E_n - \mu_I n)} .
\end{equation}
The energy density can be computed using the quantity $E_n/V$, while the isospin-charge  density can be computed from the expectation value of $n/V$, where $V$ is the volume of the system. Derivatives of observables can also be computed using 
\begin{equation}
   \frac{\partial}{\partial \mu_I} \braket{\mathcal{O}}_{\beta,\mu_I}
   = \beta \left(
   \braket{n \mathcal{O}}_{\beta,\mu_I}
   - \braket{n}_{\beta,\mu_I} \braket{\mathcal{O}}_{\beta,\mu_I}
   \right),
\end{equation}
which results from directly differentiating Eq.~\eqref{eq:obsdef}. This leads to the following expressions for the pressure, 
\begin{equation}
\label{eq:pressure}
    P(\beta,\mu_I)=\int_0^{\mu_I}\frac{\langle n\rangle_{\beta,\mu}}{V}{d \mu},
\end{equation}
and speed of sound defined by the isentropic derivative
of the pressure with respect to the energy density, $\epsilon$,
\begin{equation}
\label{eq:speed-of-sound}
    \frac{1}{c_s^2} = \frac{\partial \epsilon}{\partial P}
    = \frac{1}{\braket{n}_{\beta,\mu_I}} \frac{\partial}{\partial \mu_I} \braket{E}_{\beta,\mu_I}.
\end{equation}

Previous work~\cite{Detmold:2008fn,Detmold:2010au,Detmold:2011kw,Detmold:2012pi2,Abbott:2023coj} studied isospin-dense matter through a canonical partition function approach by using the thermodynamic relation 
\begin{equation}
    \label{eq:mu-thermo-def}
    \mu_I = \frac{dE_n}{dn}
\end{equation}
to determine the isospin chemical potential from the extracted energies. Other studies have added the isospin chemical potential directly to the QCD action~\cite{Kogut:2002tm,Kogut:2002zg,Brandt:2017oyy,Bornyakov:2021mfj,Brandt:2022hwy}. These studies probe isospin dense QCD at $\mu_I \alt 2$ and focus primarily on nonzero temperature. The results in this work are consistent with the low-temperature results in those works, but span a larger range of chemical potentials.

The primary advantage of the  method used here in comparison to the method of Ref.~\cite{Abbott:2023coj} is that $\mu_I$ enters as an input to the calculation of thermodynamic quantities rather than being derived from the isospin charge of the LQCD data and  therefore is not subject to statistical and systematic uncertainties.
A more detailed comparison to the approach of Ref.~\cite{Abbott:2023coj} is discussed in the Supplementary Material

\emph{Color-superconducting gap:}
At large isospin chemical potential, asymptotic freedom guarantees the validity of pQCD and the resulting prediction of a color-singlet superconducting state at zero temperature~\cite{son_qcd_2001}. In this state, Cooper-pairs of quark--anti-quark fields condense, leading to a superconducting gap with order parameter
$\langle \overline{d}_a\gamma_5u_b\rangle = \delta_{ab} \Delta$,
where $a$ and $b$ are color indices.
The gap $\Delta$ can be computed perturbatively \cite{son_qcd_2001,Fujimoto:2023mvc}, with the next-to-leading order
result given by
\begin{equation}
\label{eq:Delta-pert}
\Delta=\tilde{b} \mu_I \exp \left(-\frac{\pi^2+4}{16}\right) \exp \left(-\frac{3 \pi^2}{2 g}\right),
\end{equation}
where $\tilde{b}=512 \pi^4 g^{-5}$
and $g = \sqrt{4 \pi \alpha_s(\mu_I)}$ is the strong coupling at the scale $\mu_I$.
Notably, the prefactor of $1/g$ in the exponent of Eq.~\eqref{eq:Delta-pert} is smaller than the analogous coefficient in the baryon-density case  by a factor of $1/\sqrt{2}$~\cite{son_qcd_2001}, leading to an exponential enhancement of the gap and its effects in isospin-dense QCD. If pQCD is reliable for a given $\mu_I$ and $\mu_B=3\mu_I/2$, then the isospin-dense gap bounds the baryonic gap, in which there is significant phenomenological interest  \cite{Kurkela:2024xfh}.

The nontrivial background in the presence of the gap induces a change in the pressure of the system.
This change can be computed perturbatively, as has been done at NLO in Ref.~\cite{Fujimoto:2023mvc} with the result
\begin{equation}
\delta P \equiv P(\Delta)-P(\Delta=0) =\frac{N_c}{2 \pi^2} \mu_I^2 \Delta^2\left(1+\frac{g}{6}\right).
\end{equation}
This difference allows for an indirect extraction of the gap by comparing the lattice QCD pressure with the pressure derived in pQCD without pairing.

\emph{LQCD calculations}:
Following the methods and analysis techniques developed in Ref.~\cite{Abbott:2023coj}, the energies of systems of isospin charge $n\in\{1,\ldots,6144\}$ are determined from two-point correlation functions
\begin{equation}
\label{eq:correlation-function-defn}
   C_n(t) = \left\langle
    \left(\sum_{{\bf x}} \pi^-({\bf x}, 0)\right)^n \prod_{i=1}^n 
    \pi^+({\bf y}_i, t)
    \right\rangle,
\end{equation}
calculated on four ensembles of gauge field configurations whose parameters are shown in Table \ref{tab:lattice-params}.
Here, $\pi^-({\bf x},t)=\pi^+({\bf x},t)^\dagger = -\overline{d}({\bf x},t) \gamma_5 u({\bf x},t)$ is an interpolating operator built from $u$ and $d$ quark fields that creates states with the quantum numbers of the $\pi^-$.
The correlation functions are computed using the symmetric polynomial method of Ref.~\cite{Abbott:2023coj} from sparsened \cite{Detmold:2019fbk} quark propagators computed using a grid of 512 source locations on a single timeslice of each configuration. 
\begin{table*}[!t]
    \centering
\begin{tabular}{cccccccccccc}
\hline
Label & $N_\text{conf}$ & $\beta_g$ & $C_{SW}$
& $a m_{ud}$ & $a m_s$ & $(L/a)^3 \times (L_4/a)$ & $a$ (fm) & $m_\pi$ (MeV)   &$L$ (fm) & $m_\pi L$ & $T$ (MeV)\\
\hline
     A & 665 & 6.3 & 1.20537 & -0.2416 & -0.2050
    & $48^3 \times 96$ & 0.091(1) & 169(3) & 4.37 &3.75 &22.8 \\
     B & 1262 & 6.3 & 1.20537 & -0.2416 & -0.2050
    & $64^3 \times 128$ & 0.091(1) & 169(2) & 5.82 & 5.08 & 17.1\\
     C & 846 & 6.5   & 1.17008 &  -0.2091 & -0.1778
    & $72^3 \times 192$ & 0.070(1) & 164(3) & 5.04 & 4.33 & 14.7\\
     D & 977 & 6.5   & 1.17008 &  -0.2095 & -0.1793
    & $96^3 \times 192$ & 0.070(1) & 125(4) & 6.72 & 4.40  & 14.7 \\ \hline
 \end{tabular}
    \caption{Parameters of the gauge-field configurations used in this work. Ensembles were generated with $N_f=2+1$ flavors of quarks using a clover fermion action \cite{Sheikholeslami:1985ij} and a tree-level improved L\"uscher-Weisz gauge action \cite{Luscher:1984xn}.
    The first column lists the label used to refer to the ensemble, $N_{\rm conf}$ is the number of configurations, and $\beta_g$ and $C_{\rm SW}$ refer to the gauge coupling and clover coefficient, respectively. The lattice spacing, $a$, is determined in Refs.~\cite{Yoon:2016jzj,Mondal:2021oot}, while the lattice geometries are defined by the the spatial and temporal extents, $L$ and $L_4$, respectively. The bare light (up and down, $m_{ud}$) and strange ($m_s$) quark masses are given in lattice units and $m_\pi$ is the pion mass. The temperature $T=1/(aL_4)$ is also shown.}
    \label{tab:lattice-params}
\end{table*}

The relevant ground-state energies, $E_n$, are determined from an analysis of the $t$-dependence of the effective energy functions 
\begin{equation}
\label{eq:effective-mass-def}
    \begin{aligned}
   a E^\text{eff}_{n}(t) = &\log \frac{C_n(t)}{C_n(t - 1)} \\
    = &\ \vartheta_n(t) - \vartheta_n(t-1) + \frac{\sigma_n^2(t)}{2}
    -  \frac{\sigma_n^2(t-1)}{2} ,
\end{aligned}
\end{equation}
where $a$ is the lattice spacing,  $\vartheta_n(t)$ and $\sigma_n(t)$  are the mean and standard deviation of $\log C_n(t)$, and the second equality is under the assumption of log-normality \cite{Abbott:2023coj}. $N_b=2000$ bootstrap resamplings are used on each ensemble to assess the statistical uncertainties and address correlations. As in Ref.~\cite{Abbott:2023coj}, on each bootstrap the energy is given by the value of the effective mass on a randomly chosen time inside the effective mass plateau region.

Given the energies determined on each ensemble for systems of isospin charge $n\in\{1,\ldots,6144\}$, Eqs.~\eqref{eq:pressure} and \eqref{eq:speed-of-sound} are used to determine the pressure, the energy density, and speed of sound. 
The action used in these calculations is perturbatively improved, so discretization effects are ${\cal O}(a^2, g^2 a)$. The mass dependence of quantities evaluated over the range of quark masses used in the calculations is expected to be described linearly in $m_{ud}\sim m_\pi^2$. Each quantity ${X\in\{P/P_{\rm SB},\epsilon/\epsilon_{\rm SB},c_s^2/c^2\}}$ (where $P_{\rm SB} = \epsilon_{SB} / 3 = \mu_I^4 / 32 \pi^2$ is the pressure of a Stefan-Boltzmann gas) is fit with forms including arbitrary $\mu_I$ dependence and terms ${\cal O}(a^2, a^2\mu_I, a^2\mu_I^2,(m_\pi^2-\overline m_\pi^2))$,
where $\overline m_\pi =m_{\pi^+}= 139.57039$ MeV \cite{ParticleDataGroup:2022pth}, with coefficients independent of $\mu_I$. 
The systematic uncertainty from the extrapolation is assessed by combining fits with all possible subsets of the terms above through model averaging \cite{Rinaldi:2019thf,NPLQCD:2020ozd,Jay:2020jkz}. 
The systematic uncertainty and the statistical uncertainty are combined under the bootstrap procedure.
The lattice determinations of the pressure, energy density and speed of sound are shown in the Supplementary Material, along with further details.

\begin{figure}[t]
\centerline{\includegraphics[width=\linewidth]{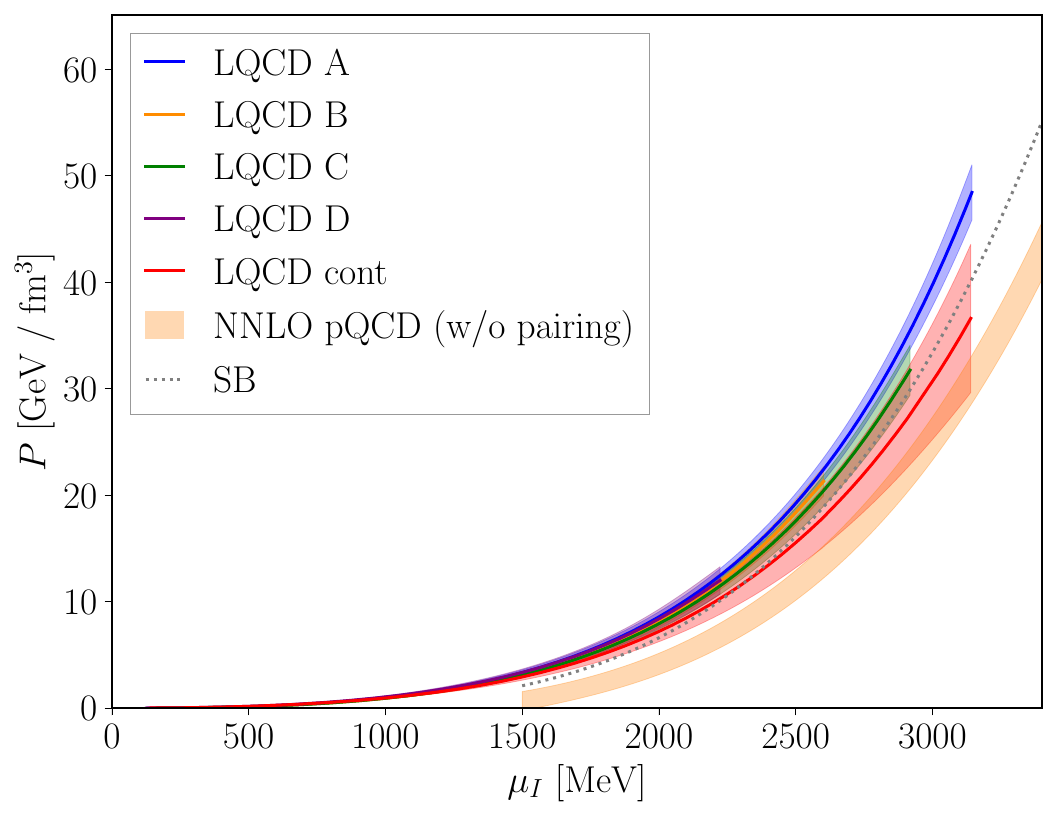}}
\caption{The pressure computed for each LQCD ensemble as well as for the continuum limit, physical mass extraction. Results from the Stefan-Boltzmann limit of a free Fermi gas and from NNLO pQCD without pairing or quark mass effects are shown for $\mu_I\agt1500$ MeV.
}
\label{fig:pressure}
\end{figure}
The calculated pressure is shown in Fig.~\ref{fig:pressure} for each lattice ensemble as well as for the continuum-limit, physical-quark-mass interpolation. The LQCD results for the different volumes, lattice spacings and quark masses used in the calculations  are seen to agree with each other within uncertainties and also with the physical mass, continuum limit extraction. The LQCD pressure agrees with NLO $\chi$PT at small values of the chemical potential and can be compared with  NNLO pQCD \cite{Freedman:1976dm,Freedman:1976ub,Freedman:1976xs,Baluni:1977ms,Kurkela:2009gj,Fujimoto:2023mvc} at large values of the chemical potential. The mild tension seen between LQCD and pQCD for $\mu_I\in[1500,2250]$ MeV potentially indicates the presence of a superconducting gap \cite{Fujimoto:2023mvc} (it could alternatively be an indication of the breakdown of pQCD, although NLO and NNLO pQCD results are in agreement over this range). 

The corresponding speed of sound is seen to exceed the conformal limit of $c_s^2/c^2=1/3$ over a wide range of the isospin chemical potential (see Fig.~\ref{fig:speed-of-sound} below). While this behavior was seen in Refs.~\cite{Brandt:2022hwy,Brandt:2022fij,Abbott:2023coj} (and also in two-color QCD \cite{Iida:2022hyy,Iida:2024irv}), the present results confirm that this is not a lattice artifact and that such behavior is possible in strongly interacting QCD matter. This suggests that the assumption that the speed of sound remains below this value in baryonic matter is questionable.

\begin{figure}[t]
\centerline{\includegraphics[width=\linewidth]{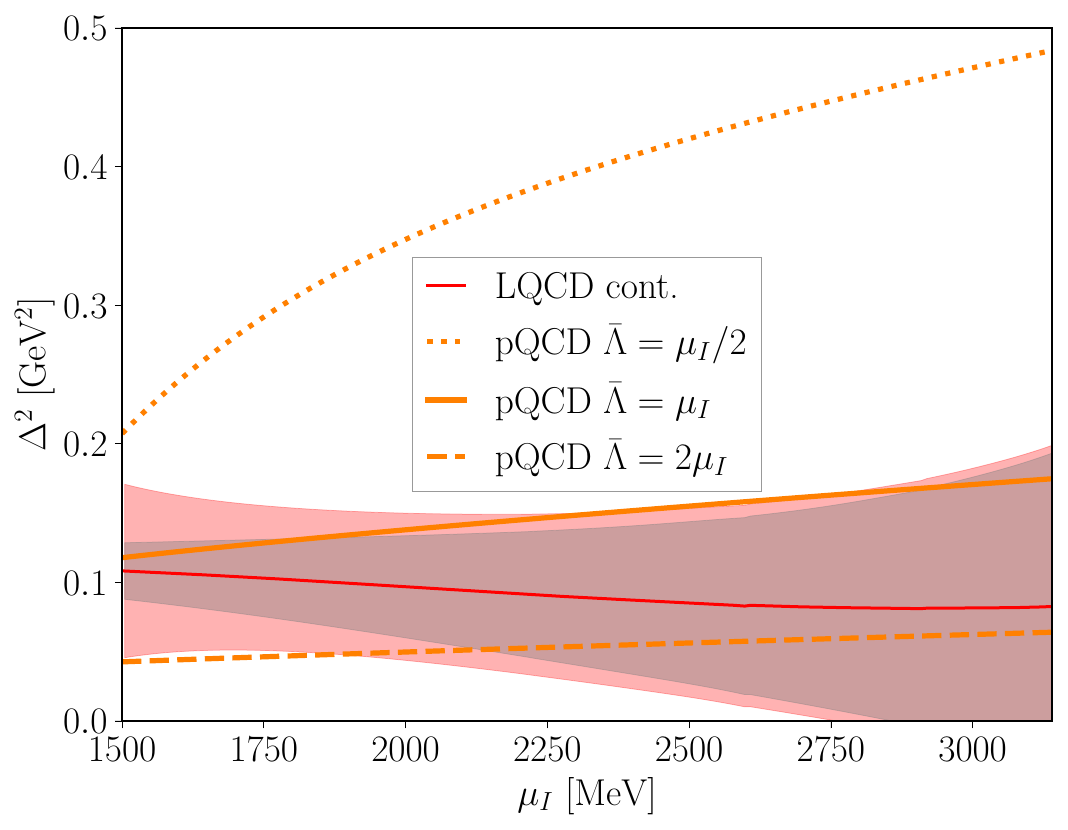}}
\caption{Comparison of the pQCD form in Eq.~\eqref{eq:Delta-pert} for the BCS gap (orange curves  evaluated at scales $\bar\Lambda=\mu_I\times\{0.5,1.0.2.0\}$) with that determined from the difference between the continuum limit of LQCD calculations and the NNLO pQCD result (red curve). The inner shaded error region is the extrapolated LQCD uncertainty while the outer shaded error region combines this with the NNLO pQCD uncertainty.
}
  \label{fig:gap}
\end{figure}

Given the LQCD calculation of the pressure, a determination of the superconducting gap can be made by subtracting the pQCD calculation of the pressure in the absence of the gap. In the range of chemical potentials where pQCD is a controlled expansion, this determines the gap, accurate to the same order as the perturbative subtraction. Figure \ref{fig:gap} shows the extracted gap found using the NNLO pQCD pressure subtraction as well as a comparison to the pQCD gap in Eq.~\eqref{eq:Delta-pert} evaluated at scales $\bar\Lambda=\mu_I\times\{0.5,1.0,2.0\}$ as a guide to uncertainty. As can be seen, the gap extracted from the LQCD calculations is in agreement with the perturbative gap for $\mu_I\in[1500,3250]$ MeV but is considerably more precisely determined than the uncertainty from perturbative scale variation.
Since there is agreement with the perturbative estimate, the gap is also most likely larger than the corresponding gap for baryonic matter.

\emph{Equation of state for isospin-dense mater:}
The continuum-limit lattice QCD calculations presented above span isospin chemical potentials from just above the pion mass to values where pQCD appears to converge. Consequently, by combining the LQCD results with $\chi$PT and pQCD, the zero temperature EoS of isospin-dense matter can be described for all $\mu_I$ with  uncertainties quantified using Bayesian inference. 
The functional dependence of each overlapping theoretical constraint on $\mu_I$ is modelled by a correlated Gaussian distribution. The ensemble of constraints is combined via a Gaussian Process (GP), following similar work for the nuclear EoS~\cite{Semposki:2024vnp,Melendez:2019izc,Essick:2020flb,Mroczek:2023zxo}. Theoretical uncertainties of $\chi$PT are estimated from the difference between the NLO and LO results, and uncertainties in pQCD are assessed from scale variation over $\bar\Lambda\in\mu_I\times[0.5,2.0]$. 
Figure \ref{fig:speed-of-sound} shows the GP-model results for the speed of sound in comparison to the three theoretical inputs.  A data file for evaluating this model accompanies this article.
With a complete quantification of the isospin-dense equation of state, phenomenological implications such as the existence of pion stars~\cite{Brandt:2018bwq,Andersen:2022aig,Stashko:2023gnn} and the isospin effects that distinguishes pure neutron matter from symmetric nuclear matter \cite{STEINER2005325,LI2008113} can be further investigated. 

\begin{figure}[t]
\centerline{\includegraphics[width=\linewidth]{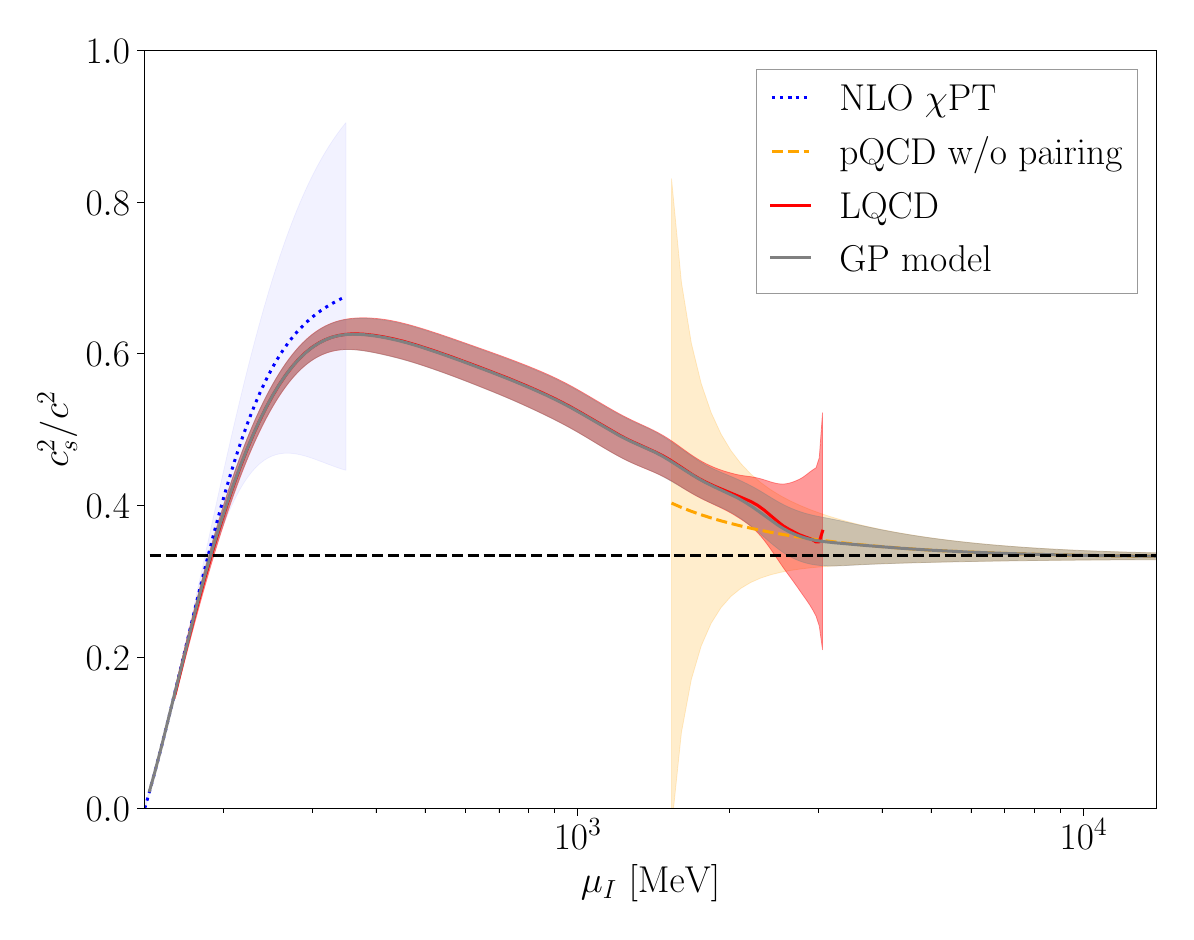}}
\caption{The squared speed of sound as a function of the isospin chemical potential. The lattice QCD determination (red), pQCD determination (orange), and $\chi$PT determination (blue) are combined into the GP model (grey) as discussed in the main text. The dashed horizontal line shows the conformal limit.
}
\label{fig:speed-of-sound}
\end{figure}

\begin{figure}[t]
\centerline{\includegraphics[width=\linewidth]{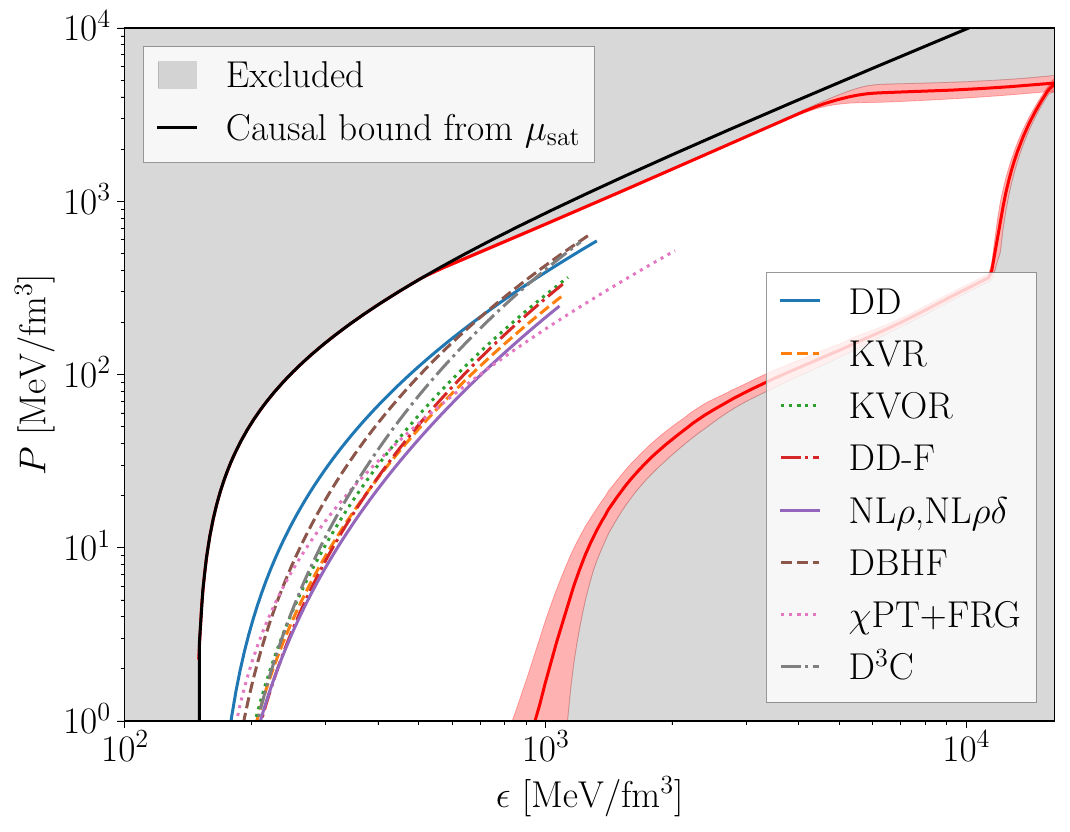}}
\caption{The bounds on the EoS of symmetric nuclear matter derived from the GP-model of the isospin-dense EoS (an EoSs that enters this region would not be consistent with QCD). The boundary of the excluded region is shown as the red band, with uncertainties propagated from the GP-model. Also shown are the bound from causality  and a range of phenomenological EoSs for symmetric nuclear matter:  DBHF \cite{vanDalen:2005sk}, DD, D$^3$C and DD-F \cite{Typel:2005ba}, KVOR and  KVR \cite{Kolomeitsev:2004ff}, NL$\rho$,NL$\rho\delta$ \cite{Liu:2001iz} (all  taken from Ref.~\cite{Leonhardt:2019fua} for $\rho_0<\rho<6\rho_0$), and a $\chi$EFT+FRG interpolation \cite{Leonhardt:2019fua} for $\rho_0<\rho<10\rho_0$. 
} 
\label{fig:nuclear-eos}
\end{figure}

\emph{Constraining the nuclear equation of state:}
The partition function of two-flavor QCD with an isospin chemical potential, $\mu_I$, can be written in terms of the path integral
 \begin{eqnarray}
Z_{\mathrm{I}}\left(\beta,\mu_{\mathrm{I}}\right)&=&
\int_\beta[d A]\operatorname{det} \mathcal{D}\left(-\frac{\mu_{\mathrm{I}}}{2}\right)\operatorname{det} \mathcal{D}\left(\frac{\mu_{\mathrm{I}}}{2}\right) e^{-S_{\mathrm{G}}}
\nonumber 
\\&=&\int_\beta[d A]\left|\operatorname{det} \mathcal{D}\left(\frac{\mu_{\mathrm{I}}}{2}\right)\right|^2 e^{-S_{\mathrm{G}}},
\end{eqnarray}
where $A$ is the gluon field, $\mathcal{D}(\mu) \equiv D\!\!\!\!/+m-\mu_{\mathrm{q}} \gamma_0$ is the Dirac operator with quark chemical potential $\mu_q$, $S_G$ is the gauge action, and  $\int_\beta[dA]$ indicates integration over gauge fields with period $\beta$ in the temporal direction. As first shown in Refs.~\cite{Cohen:2003ut,Cohen:2004qp}, this partition function
bounds the partition function of two-flavor QCD with equal chemical potentials for $u$ and $d$ quarks   
\begin{equation}
Z_{\mathrm{B}}\left(\beta,\mu_{\mathrm{B}}\right)=\int_\beta[d A] \operatorname{Re}\left[\operatorname{det} \mathcal{D}\left(\frac{\mu_{\mathrm{B}}}{N_{\mathrm{c}}}\right)\right]^2 e^{-S_{\mathrm{G}}}
\end{equation}
as
\begin{equation}
Z_{\mathrm{B}}\left(\beta,\mu_{\mathrm{B}}\right) \leq Z_{\mathrm{I}}\left(\beta,\mu_{\mathrm{I}}=\frac{2 \mu_{\mathrm{B}}}{N_{\mathrm{c}}}\right).
\end{equation}
By the monotonicity of the logarithm, the above inequality directly translates into an inequality between the pressures of the two media as a function of the energy density. 
Consequently, the isospin-dense EoS bounds the EoS for symmetric nuclear matter. At large values of the quark chemical potentials, where pQCD is valid, this bound becomes tight as differences between the partition functions enter only at ${\cal O}(\alpha_s^k)$ for $k\geq3$ \cite{Moore:2023glb}.
This bound was explored in Ref.~\cite{Fujimoto:2023unl} based on the previous lattice QCD results \cite{Abbott:2023coj} at a single lattice-spacing and unphysical quark masses. Here, Fig.~\ref{fig:nuclear-eos} presents updated bounds based on the continuum limit lattice QCD results at the physical quark masses, $\chi$PT,  and perturbative QCD through the GP-model. While the bounds from isospin-dense matter do not significantly constrain phenomenological nuclear equations of state within the uncertainties that are typically presented \cite{Annala:2023cwx,Semposki:2024vnp,Han:2022rug,Jiang:2022tps,Raaijmakers:2019dks,Huth:2020ozf,vanDalen:2005sk,Typel:2005ba,Kolomeitsev:2004ff,Liu:2001iz,Leonhardt:2019fua}, the bounds are independent of modeling uncertainties that enter the nuclear EoS in the regions that are unconstrained by nuclear structure or pQCD calculations. GP models without lattice QCD constraints result in significantly larger uncertainties in the position of the bound, in particular in the lower right corner of the red band of Fig.~\ref{fig:nuclear-eos}.

\emph{Summary:} In this letter, a determination of the equation of state of isospin-dense matter for the complete range of isospin chemical potential at zero temperature is presented for the first time. To achieve this, continuum limit LQCD calculations are combined with pQCD calculations and $\chi$PT through a model-mixing approach in overlapping regions of isospin chemical potential. Comparison to pQCD enables a determination of the superconducting gap, and QCD inequalities translate the isospin-dense EoS into rigorous bounds on the nuclear EoS relevant for astrophysical environments.

\acknowledgments{ 
We are grateful to Yuki Fujimoto and Sanjay Reddy for discussions. The ensembles used in this work were generated through the combined efforts of the JLab, William and Mary, Los Alamos, and MIT groups. We particularly thank Balint Jo\'o for assistance with the generation of the gauge configurations and quark propagators used in this work. 
The calculations were performed using an allocation from the Innovative and Novel Computational Impact on Theory and Experiment (INCITE) program using the resources of the Oak Ridge Leadership Computing Facility located in the Oak Ridge National Laboratory, which is supported by the Office of Science of the Department of Energy under Contract DE-AC05-00OR22725.
This research also used resources of the National Energy Research Scientific Computing Center (NERSC), a U.S. Department of Energy Office of Science User Facility located at Lawrence Berkeley National Laboratory, operated under Contract No. DE-AC02-05CH11231. We acknowledge USQCD computing allocations and PRACE for awarding us access to Marconi100 at CINECA, Italy.

This work is supported by the National Science Foundation under Cooperative Agreement PHY-2019786 (The NSF AI Institute for Artificial Intelligence and Fundamental Interactions, http://iaifi.org/) and by the U.S.~Department of Energy, Office of Science, Office of Nuclear Physics under grant Contract Number DE-SC0011090. 
RA, WD and PES are also supported by the U.S.~Department of Energy SciDAC5 award DE-SC0023116. RA was also partially supported by the High Energy Physics Computing Traineeship for Lattice Gauge
Theory (DE-SC0024053).
FRL acknowledges financial support from the Mauricio and Carlota Botton Fellowship. 
MI is partially supported by the Quantum Science Center (QSC), a National Quantum Information Science Research Center of the U.S. Department of Energy.
PES is also supported by the U.S. DOE Early Career Award DE-SC0021006. 
AP and RP acknowledge support from Grant CEX2019-000918-M and the project PID2020-118758GB-I00, financed by the Spanish MCIN/ AEI/10.13039/501100011033/, and from the EU STRONG-2020 project under the program H2020-INFRAIA-2018-1 grant agreement no. 824093.
This manuscript has been authored by Fermi Research Alliance, LLC under Contract No. DE-AC02-07CH11359 with the U.S. Department of Energy, Office of Science, Office of High Energy Physics. 

This work made use of Chroma \cite{Edwards:2004sx}, QDPJIT \cite{Wint1405:Framework}, QUDA \cite{Clark:2010,Clark:2016rdz}, JAX~\cite{jax2018github}, NumPy~\cite{numpy}, SciPy~\cite{2020SciPy-NMeth}, matplotlib~\cite{Hunter:2007} and HDF5 \cite{The_HDF_Group_Hierarchical_Data_Format}.
}

\bibliographystyle{utphys}
\bibliography{main}

\clearpage
\section*{Supplementary material}

\subsection{Effective mass functions and additional thermodynamic quantities}

\begin{figure*}[t]
\centerline{\includegraphics[width=\textwidth]{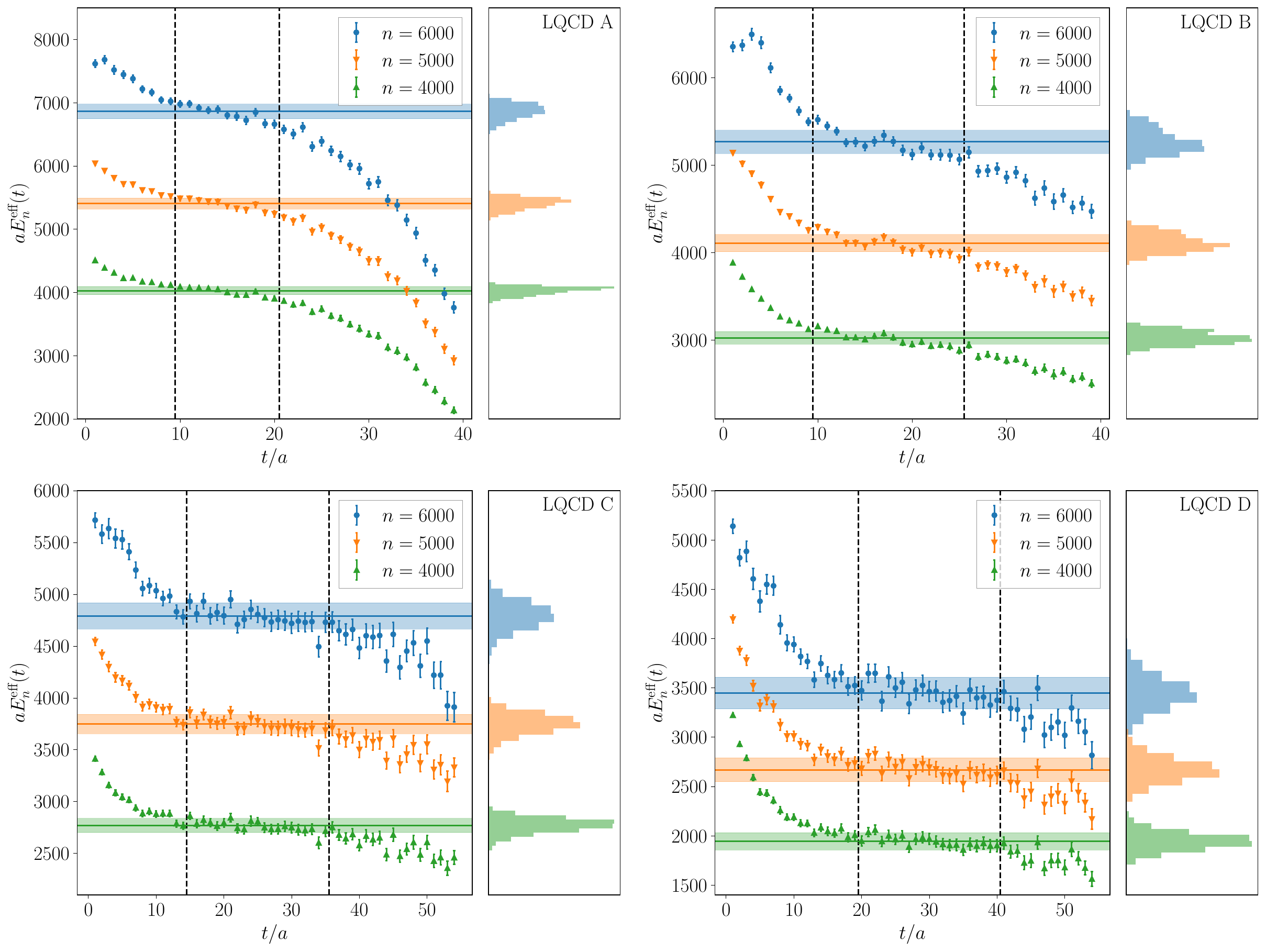}}
\caption{Effective mass functions as a function of the temporal separation for each ensemble used in this work for three different isospin charges. 
The vertical dashed lines show the time ranges used to extract the effective masses, the bootstrap histograms of which are shown at the right, as discussed in the main text. The horizontal lines and bands correspond to the mean and standard deviations across the bootstrap ensembles. }
\label{fig:effective-mass}
\end{figure*}

Figure \ref{fig:effective-mass} shows effective mass functions, Eq.~\eqref{eq:effective-mass-def}, for three different isospin charges, $n\in \{4000,5000,6000\}$, for each of the four LQCD ensembles used in this work. In all cases, the effective mass functions exhibit excited states at small times and thermal states at large times.
In order to control for these effects in the analysis, on each bootstrap sample and for each isospin charge, the effective mass is taken to be the effective mass function for a particular timeslice drawn from a uniformly random distribution of timeslices in the range  $[t_{\text{min}}, t_\text{max}]$, as in Ref.~\cite{Abbott:2023coj}. The values of $t_\text{min}$ and $t_\text{max}$ for each ensemble were chosen in order to encompass the observed plateaus in the effective mass functions for all $n$ and are shown as the vertical dashed lines in Fig.~\ref{fig:effective-mass}.

\begin{figure*}[t]
\centerline{\includegraphics[width=0.90\linewidth]{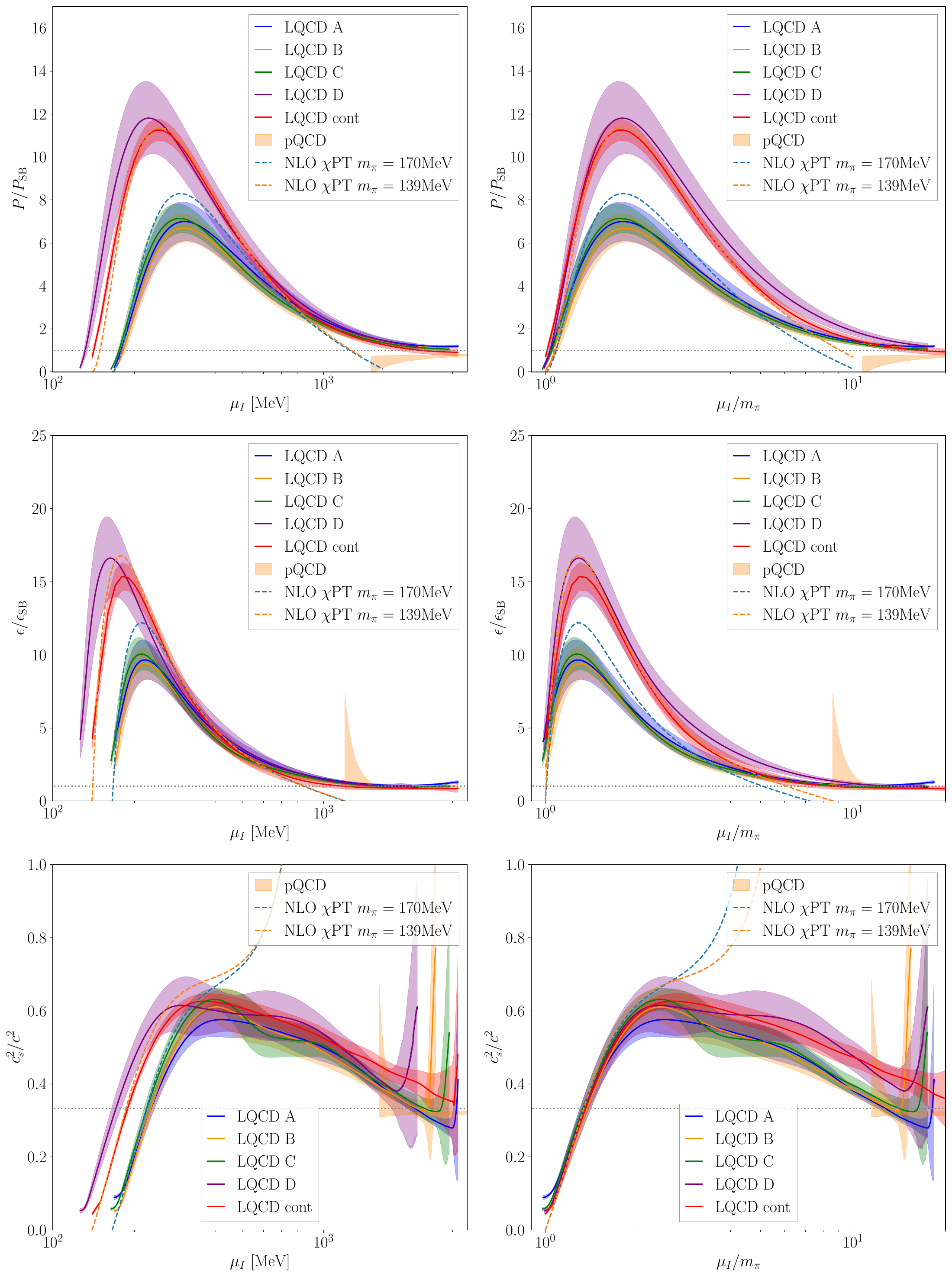}}
\caption{The normalized pressure, energy density, and squared speed of sound computed in LQCD for each ensemble as well as for the continuum limit, physical mass extraction. Results from NNLO pQCD without pairing, and NLO $\chi$PT are shown at large and small values of $\mu_I$, respectively. The horizontal gray-dotted  lines correspond to the Stefan-Boltzmann expectation for each quantity.}
\label{fig:eps-p-cs2-combined}
\end{figure*}

Figure~\ref{fig:eps-p-cs2-combined} shows two views of the normalized pressure, energy density and squared speed of sound computed on each ensemble, displaying their dependence on $\mu_I$ in MeV, and on the dimensionless ratio $\mu_I/m_\pi$. 
The threshold for the onset of nonzero pressure is at $\mu_I=m_\pi$, so the differences between the quark masses used in the various LQCD ensembles have a significant effect on thermodynamic quantities for small $\mu_I$, and the two views highlight different aspects. 
Each of the three quantities exhibits a peak at small
$\mu_I$ (at a different value for each quantity), with the location and height of the peaks varying with
the pion mass when displayed in physical units. The plots also show predictions from three-flavor NLO chiral
perturbation theory \cite{Andersen:2023ofv} at  $m_\pi = \overline{m}_\pi=139.57039$ MeV and $m_\pi=170\,\mathrm{MeV}$ (and correspondingly $m_K=493.677$  MeV and $537.6$ MeV, with a chiral cutoff scale of $\Lambda = m_\rho = 775.26$ MeV).   
The larger mass is close to the pion masses on ensembles A, B and C, while $\overline{m}_\pi$ is the physical charged pion mass and is close to the pion mass on the D ensemble.  
The variation in $\chi$PT reproduces the small $\mu_I$ differences between the LQCD results for different pion masses, indicating that
the differences exhibited at small $\mu_I$ are primarily a consequence
of the different thresholds.
For larger $\mu_I$, the results obtained on the
different ensembles agree within uncertainties.
The same quantities are also shown as a function of $\mu_I / m_\pi$ for the different ensembles used in this work, along with the continuum result. At small $\mu_I$, the results collapse onto a common curve, coinciding with the result from NLO $\chi$PT, as expected. However this normalization by the corresponding pion mass creates an artificial difference at large $\mu_I$ between the A, B and C ensembles, the D ensemble, and the continuum, physical-mass determination.
Note that the speed of sound becomes poorly determined for $\mu_I\gtrsim \mu_I^\text{max} = 22 m_\pi$; there, it has been restricted to lie in the range $[0, 1]$ as required by causality.

\subsection{Comparison with Ref.~\cite{Abbott:2023coj}}

\begin{figure}[t]
\centerline{\includegraphics[width=\linewidth]{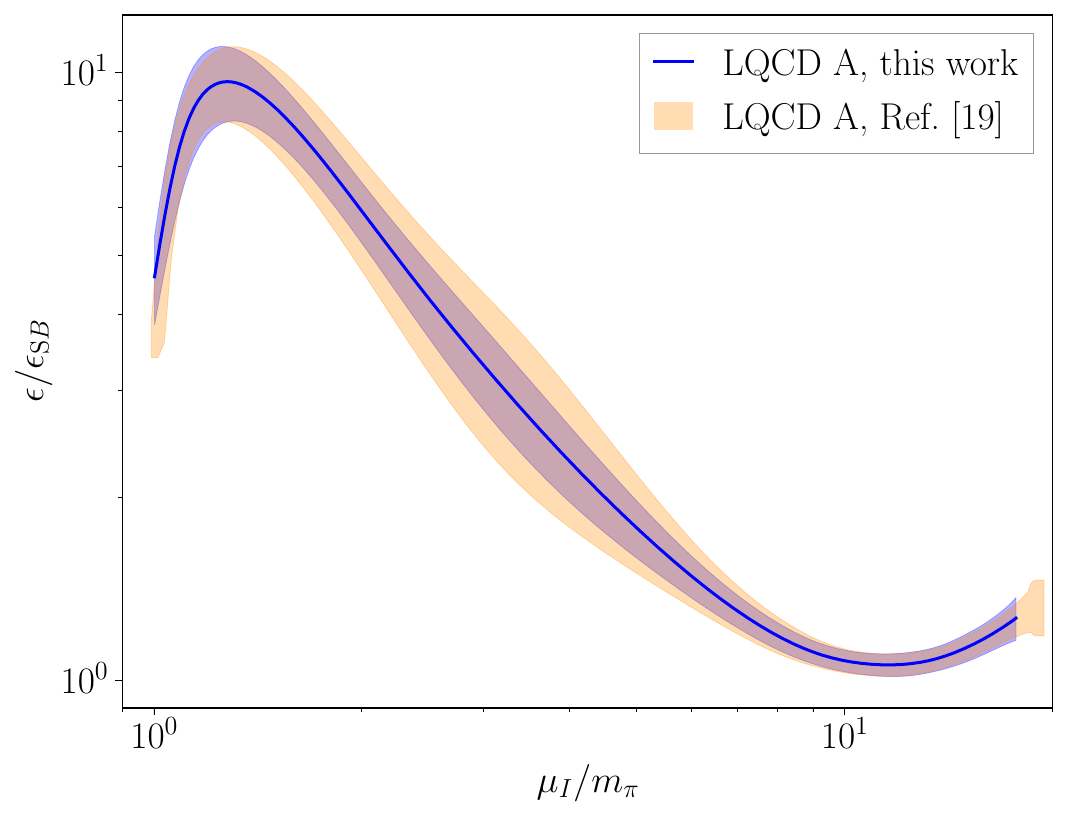}}
\caption{\label{fig:eps-compare}Comparison of the energy density ratio on the A ensemble between the method utilized in this work and the method of Ref.~\cite{Abbott:2023coj}. Uncertainties on the latter are determined using the method described in Appendix~C of of that work. }
\end{figure}
Although the methods of Ref.~\cite{Abbott:2023coj} and those used in this work  access the same underlying physics, they are not numerically identical and have different systematic uncertainties.
Figure~\ref{fig:eps-compare} shows a comparison between the energy densities obtained using the two methods.
Notably, both methods are consistent with each other, with the new method resulting in a slightly smaller uncertainty band because of the absence of an uncertainty in the determination of $\mu_I$. Similar agreement is observed for other quantities, giving confidence that both methods are equivalent within statistical uncertainties for the quantities considered in this work.

A complication absent in Ref.~\cite{Abbott:2023coj} but present in this
work is the need to determine the range of chemical potentials over which Eq.~\eqref{eq:Zdef} is a good approximation. In
Ref.~\cite{Abbott:2023coj} the chemical potential is not chosen, but is rather
derived from the thermodynamic relation in Eq.~\eqref{eq:mu-thermo-def}, while in
the current approach, the value of the chemical potential $\mu_I$ can in principle be chosen arbitrarily. In order for Eq.~\eqref{eq:Zdef} to be a valid approximation to the full
grand-canonical partition function,  $\mu_I$ needs to be chosen so that the dominant contribution to the partition function arises from $n \in \{0, \dots, n_\text{max}\}$.
The largest contribution to the grand canonical partition function arises from the $n$ that minimizes the effective energy $E_n - \mu_I n$. 
Consequently, the most relevant values of $n$ for a given value of $\mu_I$ is given by
\begin{equation}
\label{eq:n-argmax}
    n(\mu_I) = \argmin_n (E_n - \mu_I n).
\end{equation}
The minimum in Eq.~\eqref{eq:n-argmax} can be computed by setting the derivative of the argument with respect to $n$ to 0, yielding the relation
\begin{equation}
\left. \frac{dE_{n}}{dn} \right|_{n=n(\mu_I)} = \mu_I.
\end{equation}
This is exactly the thermodynamic relation given in Eq.~\eqref{eq:mu-thermo-def}, implying that a necessary and sufficient condition for a $\mu_I$ to yield sensible results is for the same value of $\mu_I$ to be attainable using thermodynamic relations. In practice, chemical potentials are used in the range $\mu_I\in[m_\pi, \mu_I^\text{max}]$,
where $\mu_I^\text{max}$ is set to be less than the value of $\left.\frac{dE_n}{dn}\right|_{n = 6000}$ for each ensemble. The exact cutoff is chosen based on statistical uncertainties.

An additional complication of the  method used in this work is the presence of the inverse
temperature $\beta$ in Eq.~\eqref{eq:Zdef}. In order to accurately represent the
zero-temperature regime ($\beta\to\infty$), $\beta$ must be taken larger than
any of the inverse energy differences in the system, while simultaneously being not so large as to induce
numerical errors from the presence of discrete energy levels.
The calculations in this work use a fixed inverse temperature $\beta = aL_4$ for
each LQCD ensemble, as listed in Table \ref{tab:lattice-params}.
For $\mu_I \agt \mu_I^{\rm min} = 1.14 m_\pi$, increasing or decreasing  $\beta$ by an order of magnitude leads to modifications of the calculated thermodynamic quantities that are orders of magnitude smaller than the corresponding statistical uncertainties. For $\mu_I < \mu_I^{\rm min}$, there is some sensitivity to $\beta$ because the system is dominated by few-pion states, and thus is far from the thermodynamic limit. Consequently, LQCD data with $\mu_I < \mu_I^{\rm min}$ are excluded from any further analysis.

\subsection{Continuum/chiral extrapolation details}
\label{app:extrap}

The continuum limit and the physical quark mass interpolation are achieved in this work using a correlated fit of the LQCD data for $X \in \{P/P_{SB}, \epsilon / \epsilon_{SB}, c_s^2/c^2\}$ evaluated at $N_\mu=300$ values, $\{\mu_I^{(1)}, \ldots,\mu_I^{(N_\mu)}\}$. The fits are performed for each bootstrap sample\footnote{Here, the $j$th bootstrap sample refers to the combination of the same bootstrap element from each of the four LQCD datasets.}, $j\in \{1,\ldots,N_b\}$, using the form
\begin{eqnarray}
\label{eq:extrap}
    X^{(j)}(a,\mu_I^{(m)},m_\pi) &=& X_0^{(j)}(\mu_I^{(m)},m_\pi)
    + X_1^{(j)}a^2
   \nonumber \\ &&
     + X_2^{(j)}a^2\mu_I^{(m)} 
    +X_3^{(j)}a^2(\mu_I^{(m)})^2 
       \nonumber \\ &&
 +X_4^{(j)}\Delta m_\pi^2, \quad \forall m\in\{1,\ldots,N_\mu\}
 \nonumber \\
 \end{eqnarray}
where $\Delta m_\pi^2=m_\pi^2-\overline m_\pi^2$. In this fit, the  $X_0^{(j)}(\mu_I^{(m)},\overline{m}_\pi)$ are the continuum, physical-pion-mass values for $X$ on bootstrap sample $j$ at each of the given $\mu_I^{(m)}$, and $X_1^{(j)}, \dots, X_4^{(j)}$ are $\mu_I$ independent coefficients. 
The arbitrary function $X_0^{(j)}(\mu_I,m_\pi)$ must accommodate two competing effects as discussed above: a) the presence
of sharp mass-dependent features in thermodynamic quantities due to
the phase transition at $\mu_I = m_\pi$, and b) the expected mass independence for $\mu_I\gg m_\pi$. 
In order to provide a
 chiral-continuum extrapolation that smoothly transitions between these two regimes, $X_0^{(j)}(\mu_I,m_\pi)$ is parameterized as 
\begin{equation}
    X_0^{(j)}(\mu_I,m_\pi)=X_0^{(j)}(\mu_I^{\rm phys})
\end{equation}
where the effective chemical potential $\mu_I^{\rm phys}$ is determined from the relation
\begin{equation}
\label{eq:muIphys}
    \mu_I =\left(1 + \left[ \frac{m_\pi}{\overline{m}_\pi} - 1 \right] \frac{1}{1 + e^{2(\mu_I^\text{phys}/\overline{m}_\pi - \alpha)}} \right) \mu_I^\text{phys},
\end{equation}
with  $\alpha\in \mathbb{R}$. Note that as $\mu_I \to m_\pi$, the effective chemical potential
satisfies $\mu_I^{\text{phys}} / \overline{m}_\pi \approx \mu_I /
m_{\pi}$, while $\mu_I^\text{phys} \approx \mu_I$ for $\mu_I \gg
m_\pi$. Furthermore,  the parameterization is trivial at the physical quark masses where
$\mu_I^\text{phys} = \mu_I$. The value of $\alpha$ is allowed to vary over the bootstrap samples,
chosen uniformly from the range $\alpha \in[2.5, 5.5)$. The transformation used in Eq.~\eqref{eq:muIphys} is certainly not unique and the variation of $\alpha$ builds in an uncertainty associated with the choice of parameterization.\footnote{Note that this parameterization is used only for computing the continuum, physical-pion-mass results, and results for the individual LQCD ensembles are displayed in terms of the unmodified chemical potential $\mu_I$.} Residual quark mass dependence not associated with the phase transition and asymptotic behaviour for $\mu_I\gg m_\pi$ is expected to be smooth and is parameterized by the last term in Eq.~\eqref{eq:extrap}. 

Note that the fit form in Eq.~\eqref{eq:extrap} does not include an infinite volume extrapolation because no statistically-significant volume dependence is observed in the LQCD data except for $\mu_I\sim m_\pi$ where the system is not in the thermodynamic limit (data with $\mu_I< \mu_I^{\rm min}$ are excluded from further analysis and not used as constraints in the GP model).
Fits capture the correlations between LQCD data at different $\mu_I$ through a covariance matrix 
$\Sigma \mapsto s \Sigma + (1 - s) \Sigma_\text{uncorr}$ applied with shrinkage parameter $s = 0.9$ for numeric stability (here $\Sigma$ denotes the covariance matrix of the data, and $\Sigma_\text{uncorr}$ is the uncorrelated covariance matrix, i.e., the diagonal entries of $\Sigma$).

\begin{figure*}[t]
\centerline{
\subfloat[]{\includegraphics[width=0.48\linewidth]{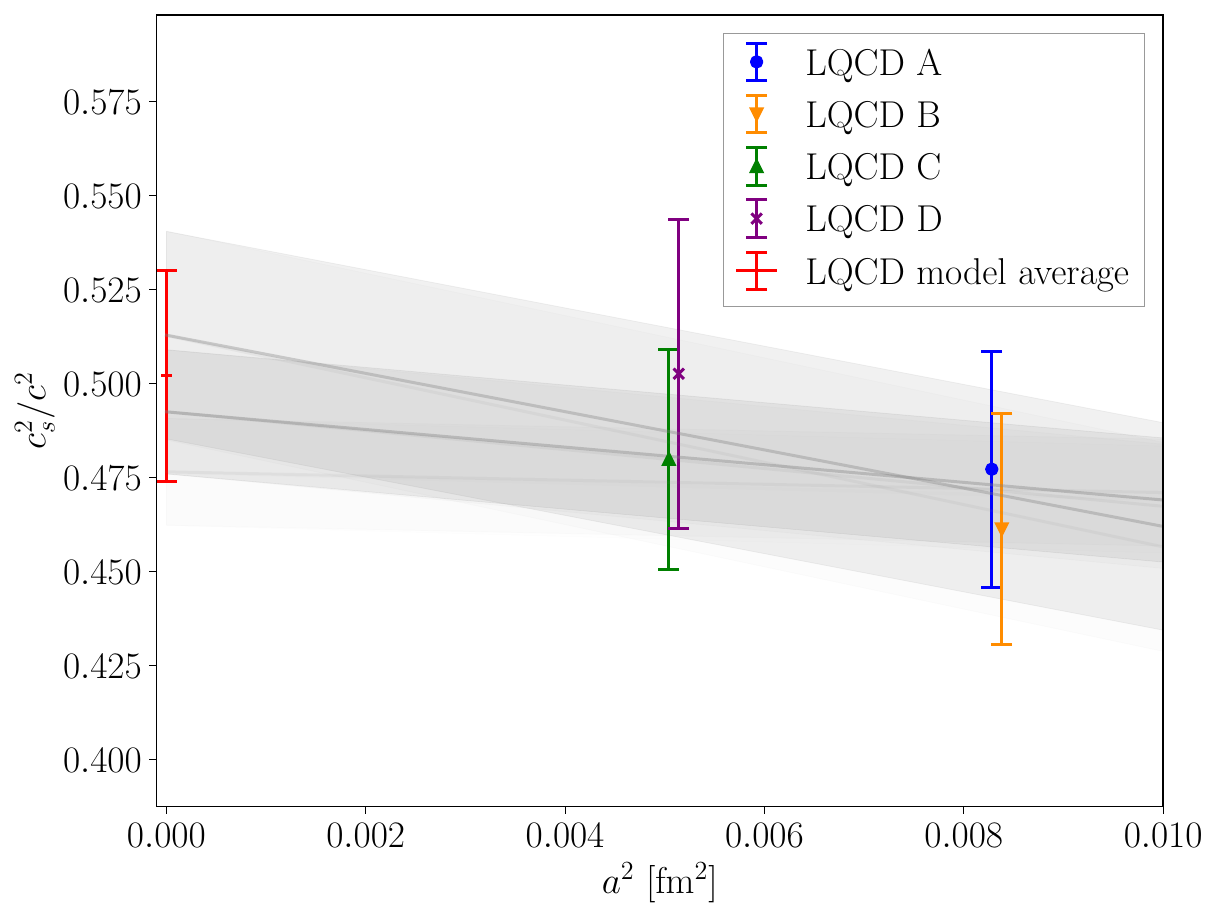}} \qquad
\subfloat[]{\includegraphics[width=0.48\linewidth]{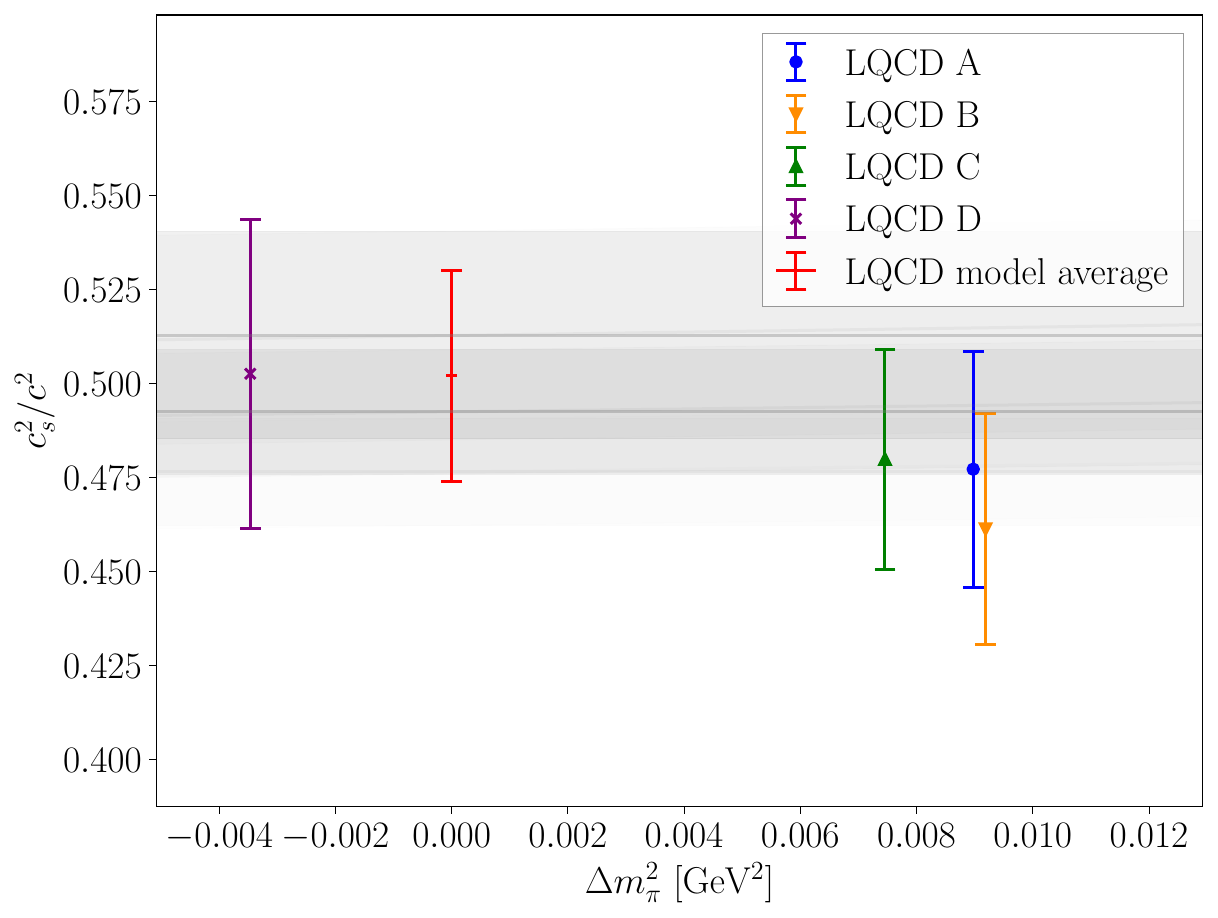}}
}
\caption{\label{fig:continuum-extrap}Continuum (a) and chiral (b) projections of the fits to the squared speed of sound evaluated at $\mu_I=1143$  MeV. The grey bands in each panel correspond to all 16 fits that enter the model average, with opacity determined by the AIC model-weight of the given fit. The model average in the continuum and at the physical quark mass is shown as the red data point in each panel. LQCD data points with the same lattice spacing or quark mass are offset slightly for clarity.}
\end{figure*}

\begin{figure}[t]
\centerline{\includegraphics[width=0.99\columnwidth]{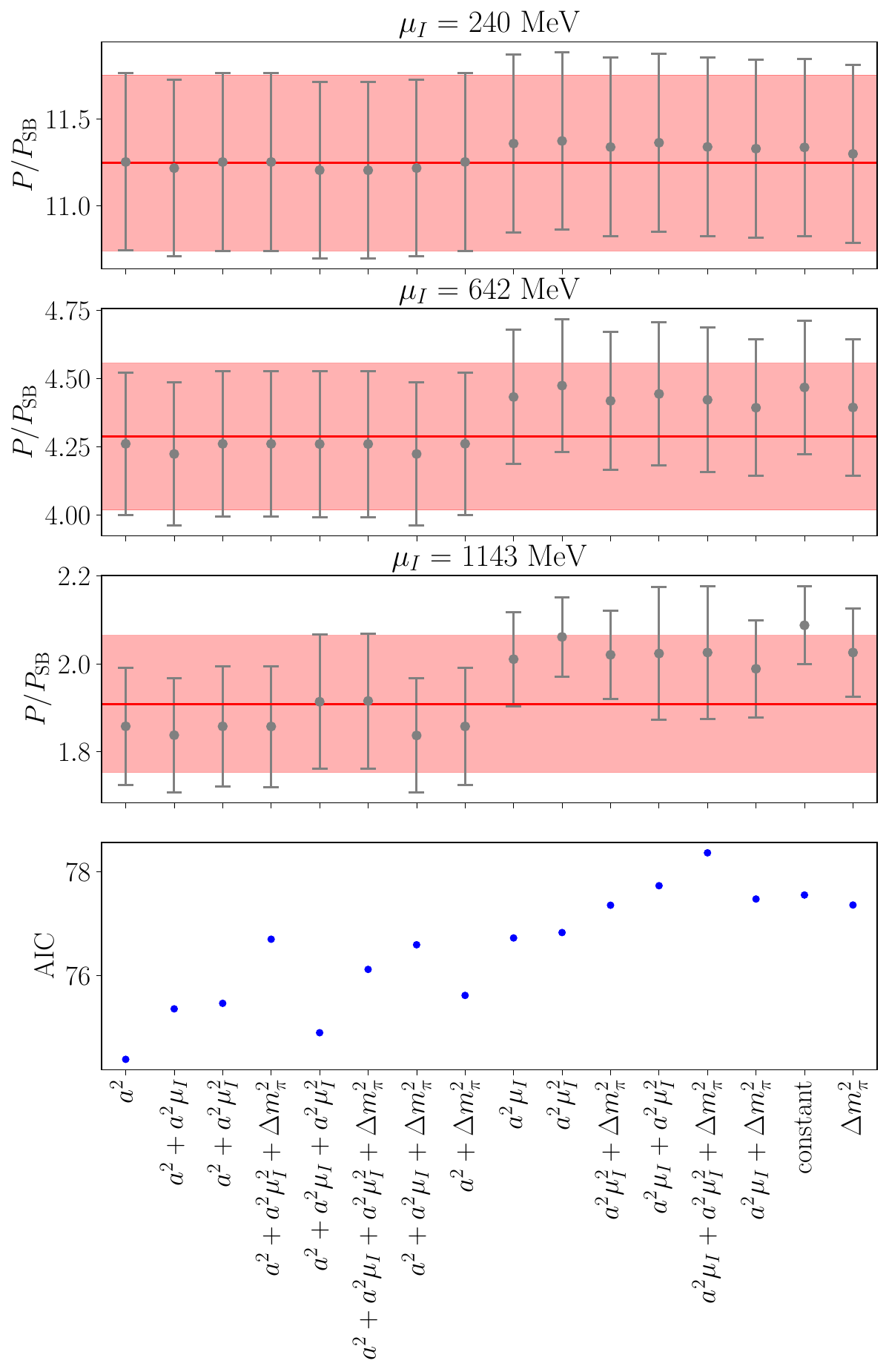}}
\caption{Values of the pressure ratio $P / P_{\rm SB}$ for all of the continuum and chiral extrapolation models considered in this work, evaluated at $\mu_I\in\{240,642,1143\}$ MeV, along with the average value of the AIC for each model. The red lines and bands indicate the model-averaged result and its uncertainty at each given chemical potential.
}
\label{fig:aic}
\end{figure}

A separate fit is performed using each possible combination of lattice spacing and mass dependent terms in Eq.~\eqref{eq:extrap}, and the results are combined using the Bayesian model averaging approach described in Ref.~\cite{Jay:2020jkz}, which assigns a weight to each fit of the form
\begin{equation}
  e^{-\rm{AIC}} = e^{-(\chi^2 + 2k)},
\end{equation}
where $\rm{AIC}$ is the Akaike information criterion, $\chi^2$ indicates the
correlated $\chi^2$ of the fit, and $k$ is the number of free parameters of the
fit. Example fits are shown in Fig.~\ref{fig:continuum-extrap},
and the averaged values of the AIC are shown in Fig.~\ref{fig:aic}, along with the resulting fits for all models at three values of the chemical potential. The
smallest AIC is achieved by the fit that contains only
$a^2$ (i.e., taking $X_2^{(j)} = X_3^{(j)} = X_4^{(j)} = 0$ in
Eq.~\eqref{eq:extrap}), although several other fits have similar
AIC values resulting in a nontrivial model
average. Note that a separate model averaging procedure is performed
 for each bootstrap sample, so the averaged AIC values in
Fig.~\ref{fig:aic} are only representative and individual bootstrap
samples will have a different distribution of weights across the
various fits.

Since each thermodynamic quantity considered within this work is sampled at $N_\mu$ points on each of the four ensembles, there are a total of $4N_\mu=1200$ data values to fit. Comparing with the values of the AIC in Fig.~\ref{fig:aic}, this indicates that the net $\chi^2$ per degree of freedom is approximately $65 / 1200 \sim 0.05$, a rather low value that  naively indicates severe over-fitting of the data even for a constant extrapolation. In this context, however, the $\chi^2$ per degree of freedom is  artificially small because of the choice to over-sample the chemical potential. For the chosen value of $N_\mu$, thermodynamic quantities at adjacent values of $\mu_I$ are strongly correlated and hence should not be considered as independent degrees of freedom. Instead, a better methodology for counting the number of degrees of freedom is to use the width of the data measured in correlation lengths across $\mu_I$. Empirically, this width corresponds to $\sim10$ degrees of freedom per ensemble (see Appendix~\ref{sec:model-mixing}), leading to a total of $\sim 40$  degrees of freedom across all ensembles, which results in a $\chi^2$ per effective degree of freedom of ${\cal O}(1)$.

\subsection{Further details on the Gaussian Process model}\label{sec:model-mixing}

\subsubsection{Gaussian Process}

Here, the procedure to compute a Bayesian model mixture for the thermodynamic quantities is described, following the approach used for the EoS at nonzero baryon chemical potential in Ref.~\cite{Semposki:2024vnp}. With this approach, it is possible to combine three different theoretical constraints (pQCD, LQCD and $\chi$PT) that take the form of a functional dependence on the isospin chemical potential $\mu_I$ using a GP. In order to explain the method, a single thermodynamic quantity $X$ is first considered before the method is extended to the case of several correlated quantities. 

In this procedure, one assumes that the underlying function $X(x \equiv \log_{10} \mu_I / \overline{m}_\pi )$, such as the pressure, follows a GP. Note that the functions are parameterized in the variable $\log_{10} \mu_I/\overline{m}_\pi$ rather than $\mu_I$  because $\mu_I$ varies over several orders of magnitude. A prior of the form
\begin{eqnarray}
    X(x) \sim {\rm GP}[f_p(x), K_p (x,x')]
\end{eqnarray}
is adopted, where $f_p(x)$ is the functional form of the prior, and $K_p (x,x')$ represents the covariance between two points of evaluation of the function (note that the subscript $p$ refers to the prior). A standard choice~\cite{Semposki:2024vnp} is:
\begin{eqnarray}
    K_p (x,x') = \sigma_p(x) \sigma_p(x') \kappa(x,x', \ell_p) + \epsilon_p \delta_{x,x'},
\end{eqnarray}
where $ \sigma(x)$ is the uncertainty at a given value of $\log_{10} \mu_I/\overline{m}_\pi$, $\ell_p$ is a hyperparameter that plays the role of a correlation length, and $\kappa$ is a kernel which is conventionally chosen to be 
\begin{eqnarray}
\kappa(x, x', \ell) = \exp \left[  -\frac{(x-x')^2}{2\ell^2} \right].    
\label{eq:kernel}
\end{eqnarray}
It is useful to include a small $\epsilon_p$ in the diagonal of the covariance matrix for numerical stability.

Each theoretical constraint is also modelled by a GP of the form:
\begin{eqnarray}
    Y_i = {\rm GP}[f_i(x), K_i (x,x', \ell_i)],
\end{eqnarray}
where $i\in\{\chi{\rm PT},{\rm LQCD},{\rm pQCD}\}$ labels the constraint and
\begin{eqnarray}
    K_i (x,x') = \sigma_i(x) \sigma_i(x') \kappa(x,x', \ell_i) + \epsilon_i \delta_{x,x'},
\end{eqnarray}
in the corresponding range of applicability of the constraint,
\begin{equation}
 x^{{\rm min}}_i <   x < x^{{\rm max}}_i.
\end{equation}

In order to combine all the available information from the theoretical constraints, a set of $N_{\rm eval}$ evaluation points in the range $[x^{\rm min}, x^{\rm max}]$ are chosen, denoted as $\{x_k\}$.\footnote{Note that in this application, all three constraints can be evaluated at arbitrary points within their intervals. Compared to Ref.~\cite{Semposki:2024vnp}, no differentiation between ``training'' and ``evaluation'' points is required here. }
The prior and the constraints are combined over these evaluation points in a Bayesian manner to produce the posterior probability distribution:
\begin{align}
\begin{split}
    \log p( f(\{x_k\}) | f_i, K_i  ) &=
    \log p(f_p(\{x_k\}), K_p ) \\&+ \sum_i \log  p(f_i(\{x_k\}), K_i ).
\end{split}
\end{align}
Since all constraints take the form of a multivariate Gaussian distribution, the posterior is also Gaussian and has inverse covariance:
\begin{equation}
    K^{-1} = K^{-1}_p + \sum_i P_i K^{-1}_i P_i,
\end{equation}
where $P_i$ is a projector that ensures that only contributions from within the validity interval of each constraint are used. That is, $P_i$ is a diagonal matrix with 
\begin{equation} 
(P_i)_{jk} =\left\{\begin{array}{ll}
\delta_{jk} & {\rm if}\ x_k \in [ x^{{\rm min}}_i, x^{{\rm max}}_i] \\ 
0 & {\rm otherwise} \end{array}\right..
\end{equation}
Similarly, the mean of the posterior is: 
\begin{equation}
    y = K \left( K^{-1}_p y_p + \sum_i P_i K^{-1}_i P_i y_i  \right).
\end{equation}
Samples of the GP model over the evaluation points $\{x_k\}$ can be thus obtained by sampling from a multivariate Gaussian distribution with mean $y$ and covariance $K$. If a continuous function is needed, one can interpolate using cubic splines, for example.

The previous construction holds for a single thermodynamic quantity, however in this work, a model mixture is required for three correlated quantities: $P/P_{\rm SB}$, $\varepsilon/\varepsilon_{\rm SB}$ and $c_s^2/c^2$. The generalization to this case is straightforward: all quantities are considered to follow a GP and the covariance between different quantities must also be modeled. For instance, the prior becomes
\begin{eqnarray}
    \vec X(x) \sim {\rm GP}[\vec f_p(x), K^{(nm)}_p (x ,x')],
\end{eqnarray}
where $\vec X$ is a vector containing all three quantities to consider, and the covariance is modelled as:
\begin{align}
    \begin{split}
     K^{(nm)}_p (x ,x') =&r_p^{(nm)} \sigma^{(n)}_p(x)  \sigma^{(m)}_p(x')  \kappa(x,x', \ell^{(nm)}_p) \\
     & +\epsilon_p \delta_{n,m}\delta_{x,x^\prime},
    \end{split}
\end{align}
where $n$ and $m$ are indices that indicate components of $\vec X$, $r_p^{(nn)}=1$, $r_p^{(n, m \neq n)}$ indicates correlation between variables, and $ \sigma^{(n)}_p(x)$ is the uncertainty on the quantity $X^{(n)}$. Similar generalizations to the case of several correlated quantities hold for the theoretical constraints $\vec Y$ and their covariances, $K^{(nm)}_i$ for $i\in\{\chi{\rm PT},{\rm LQCD},{\rm pQCD}\}$.

\subsubsection{ Description of the GP model }

The GP model presented in the main text has several Bayesian choices in the hyperparameters that are listed here. The quantities are ordered as $\vec X = \left\{ c_s^2/c^2, P/P_{\rm SB}, \epsilon/\epsilon_{\rm SB} \right\}$. In all cases, the covariance matrix is regulated with $\epsilon_p = \epsilon_{\chi{\rm PT}} = \epsilon_{{\rm LQCD}} = \epsilon_{{\rm pQCD}} = 10^{-5}$. The number of evaluations, ${N_{\rm eval} = 50}$, is chosen such that increasing that number leads to changes in the results that are much smaller than the uncertainty of the GP model.

\begin{table}
\begin{equation*} 
\begin{aligned}
\vec f_p (x) &= {\footnotesize \begin{pmatrix}
        1/3 \\ 7 \\ 10
\end{pmatrix}} \\
\vec \sigma_p (x) &= {\footnotesize \begin{pmatrix}
        1/3 \\ 7 \\ 10
\end{pmatrix}} \\
\ell_p= \ell_{\chi{\rm PT}} = \ell_{\rm pQCD}&=  {\footnotesize \begin{pmatrix}
       0.05 & 0.05 & 0.05 \\
         0.05 & 0.05 & 0.05 \\
          0.05 & 0.05 & 0.05
\end{pmatrix}} \\
 r_p=r_{\chi{\rm PT}}= r_{\rm pQCD}&= {\footnotesize \begin{pmatrix}
       1 & 0.1 & 0.1 \\
         0.1 & 1 & 0.1 \\
          0.1 & 0.1 & 1 
\end{pmatrix}}\\
 \ell_{\rm LQCD} &= {\footnotesize \begin{pmatrix}
        0.1 & 0.1 & 0.1 \\
         0.1 & 0.3 & 0.1 \\
          0.1 & 0.1 & 0.2 
\end{pmatrix}} \\
r_{\rm LQCD} &= {\footnotesize \begin{pmatrix}
       1 & -0.05 & -0.05 \\
         -0.05 & 1 & 0.4 \\
          -0.05 & 0.4 & 1 
\end{pmatrix}} \\
[x_{\chi{\rm PT}}^{\rm min},x_{\chi{\rm PT}}^{\rm max}] &= [0, \log_{10} 2.5] \\
[x_{\rm pQCD}^{\rm min},x_{\rm pQCD}^{\rm max}] &=
[\log_{10} 11 ,\infty]  \\ 
 [x_{\rm LQCD}^{\rm min},x_{\rm LQCD}^{\rm max}] &= [\log_{10} 1.14,\log_{10} 22] 
\end{aligned}
\end{equation*}
\caption{GP-model parameters.
\label{tab:GPmodel-parameters}
}
\end{table}

First, the functional form of the prior is chosen to be a constant whose value is similar to the LQCD data, and the prior width is set  so that it does not impact the results. 
The corresponding hyperparameters are listed in 
Table~\ref{tab:GPmodel-parameters}.

For the $\chi$PT constraint, the central values for each quantity, $\vec{f}_{\chi{\rm PT}}(x)$, are chosen to be the NLO expressions from Ref.~\cite{Adhikari:2019zaj} using the LECs provided in Table X of that reference. The uncertainties $\vec\sigma_{\chi{\rm PT}}(x)$  for the three quantities are chosen to be the absolute value of the difference between the LO and NLO calculations. 

For the pQCD constraints, the components of $\vec{f}_{{\rm pQCD}}(x)$ are chosen to be the NNLO expression in the massless limit from Ref.~\cite{Kurkela:2009gj} including the effects of the superconducting gap at NLO~\cite{Fujimoto:2023mvc}.
The uncertainties $\vec\sigma_{{\rm pQCD}}(x)$ are  chosen to be half of the difference of the scale variation of each quantity in the interval $\Lambda \in \mu_I\times[0.5,2.0]$.  The correlation length and the correlations between different thermodynamic quantities in both $\chi$PT and pQCD is also chosen to be the same as in the prior. The assumed range of validity of both theoretical description is given in Table~\ref{tab:GPmodel-parameters},
along with the other hyperparameters. 

For the LQCD constraint, the central values are the model-averaged values after the continuum chiral extrapolation described in Sec.~\ref{app:extrap}. 
The uncertainties $\vec\sigma_{{\rm LQCD}}(x)$ are evaluated from the bootstrap samples, i.e., the diagonal part of the covariance matrix is obtained through bootstrap. 
The off-diagonal entries of the covariance matrix are modeled to follow the kernel in Eq.~\eqref{eq:kernel}, with values that represent the empirical correlations in the bootstrap samples. The range of validity is such that it avoids regions at very small and very large chemical potential where the LQCD calculations have larger uncertainties, $\mu_I^{{\rm min}}<\mu_I<\mu_I^{{\rm max}}$, as discussed above. All relevant quantities are listed in Table~\ref{tab:GPmodel-parameters}.

\subsubsection{Additional thermodynamics quantities from the GP model}

\begin{figure}[ht!]
     \centering
     \subfloat[\label{fig:prat-mixture}]{
     \includegraphics[width=0.92\columnwidth]{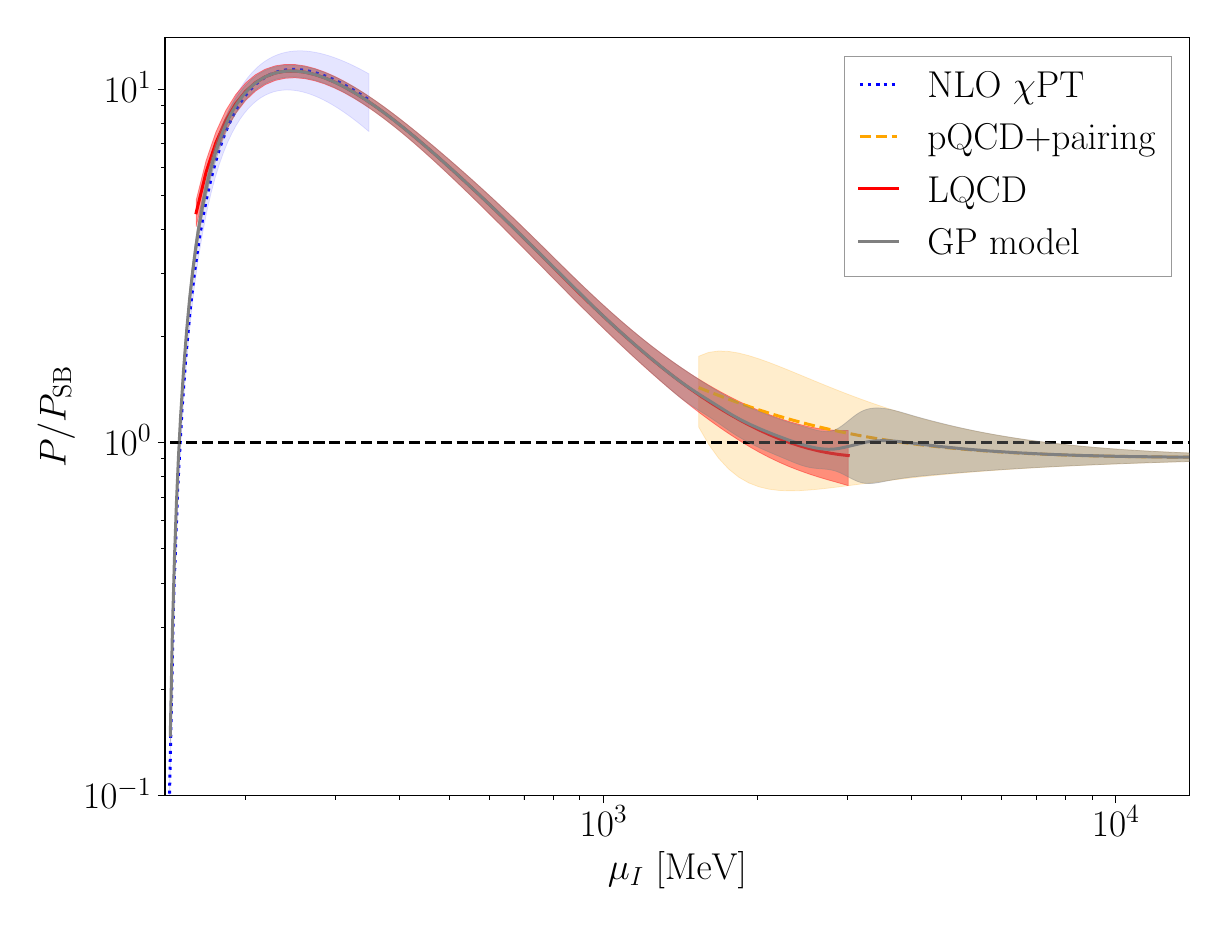}
    } \\
    \subfloat[\label{fig:edens-mixture}]{
     \includegraphics[width=0.92\columnwidth]{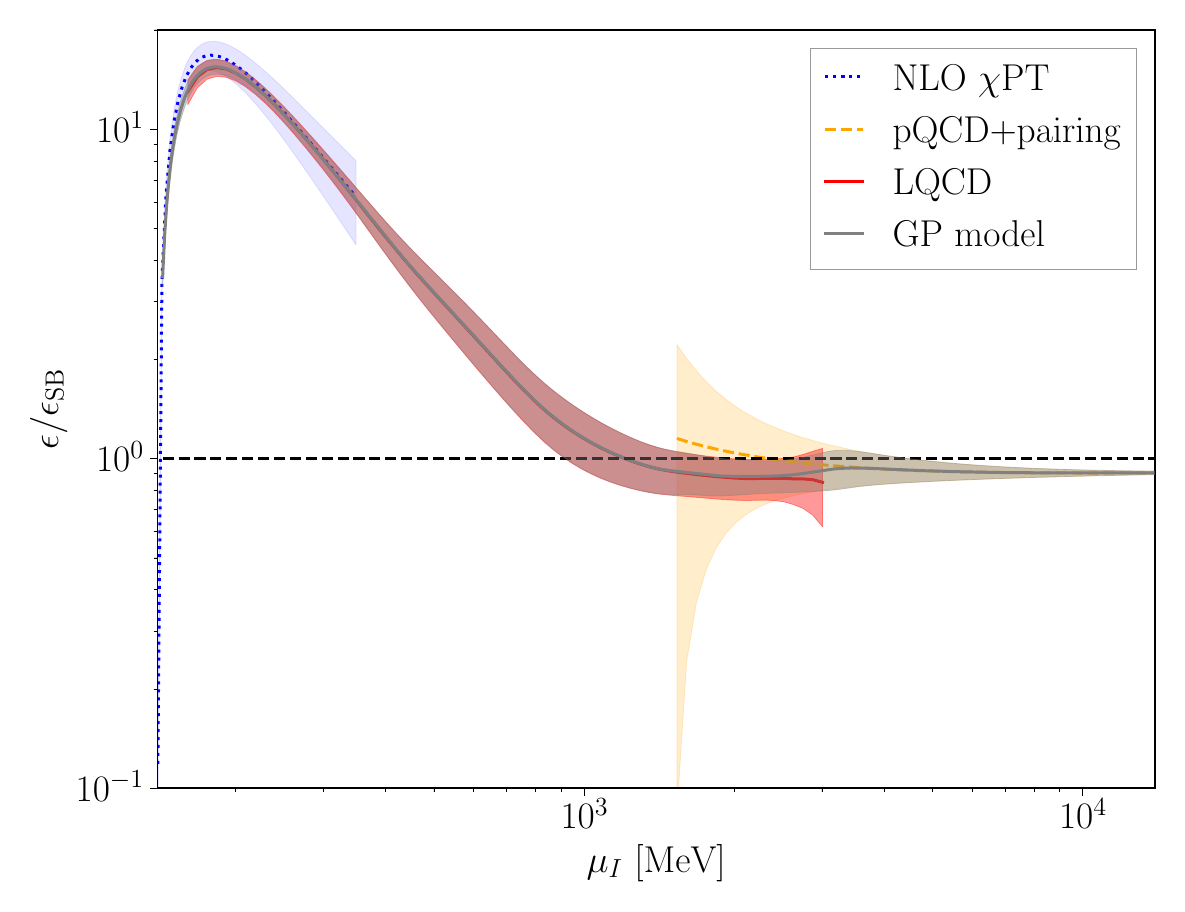}
    }
    \caption{ Same as Fig.~\ref{fig:speed-of-sound}, but for the pressure ratio (a) and energy density ratio (b) to the Stefan-Boltzmann result. } 
    
\end{figure}

The squared speed of sound for the GP model is shown in Fig.~\ref{fig:speed-of-sound} of the main text. Figs.~\ref{fig:prat-mixture} and \ref{fig:edens-mixture} show the corresponding pressure and energy density ratios to the Stefan-Boltzmann (free Fermi gas) expectation.

\begin{figure}[ht!]
     \centering
      \subfloat[]{
     \includegraphics[width=0.92\columnwidth]{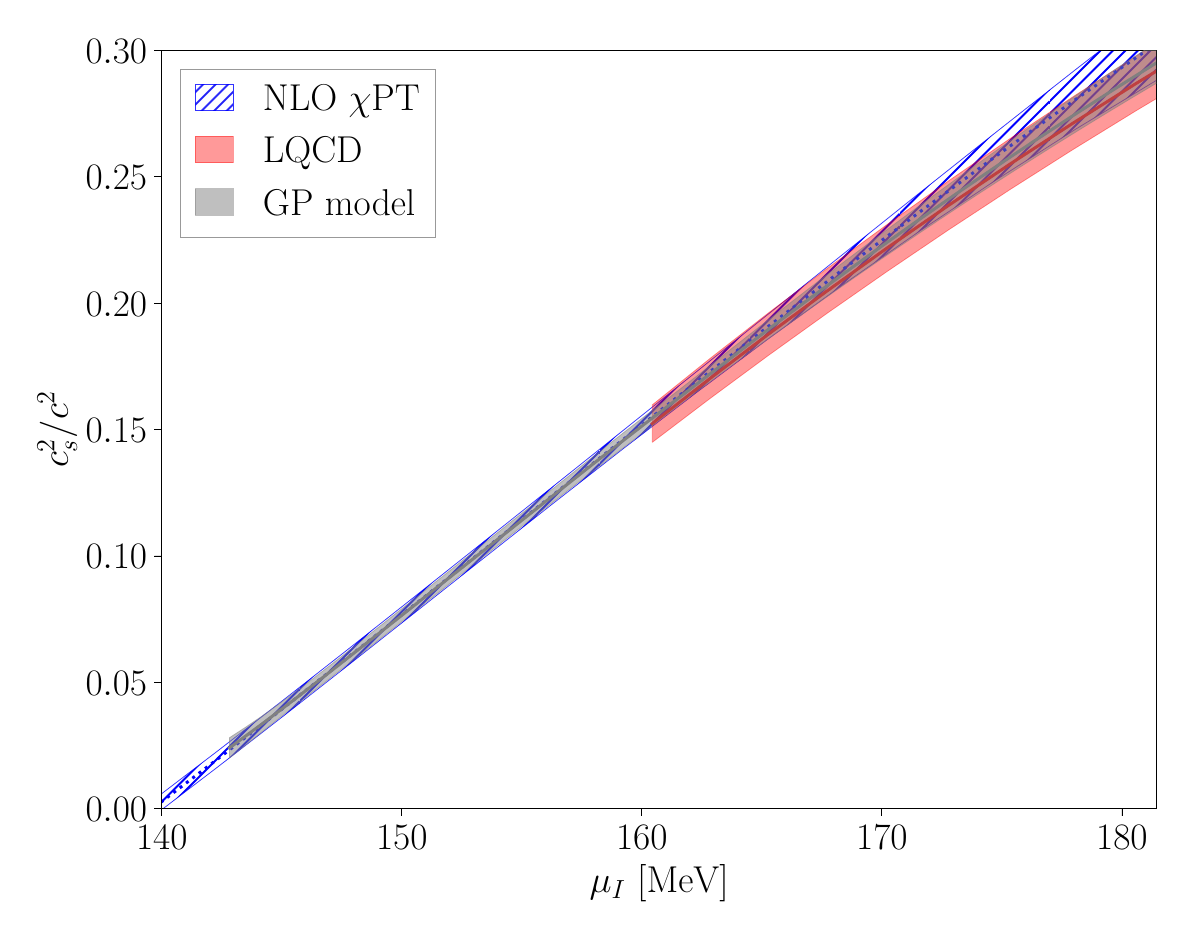}
    } \\
     \subfloat[]{
     \includegraphics[width=0.92\columnwidth]{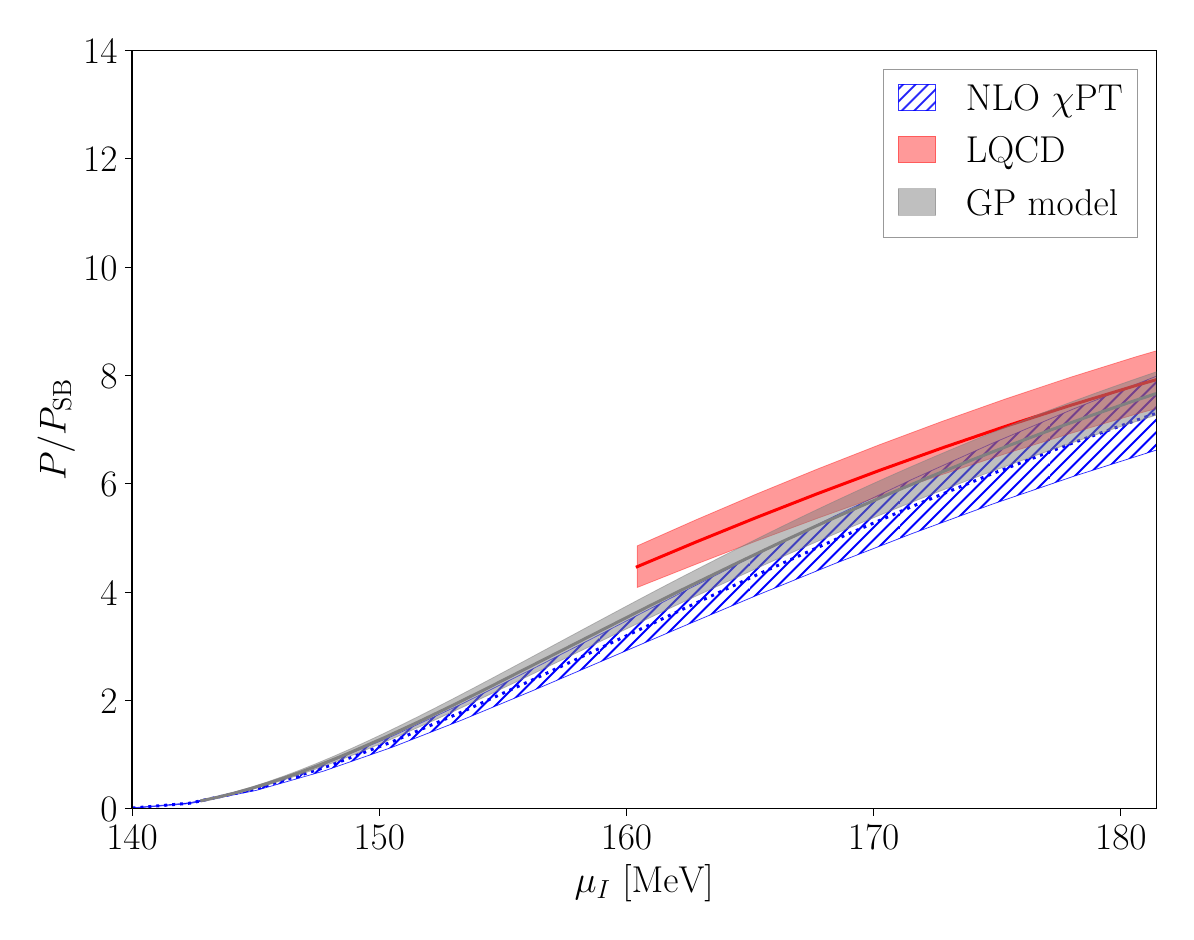}
    } \\
    \subfloat[]{
     \includegraphics[width=0.92\columnwidth]{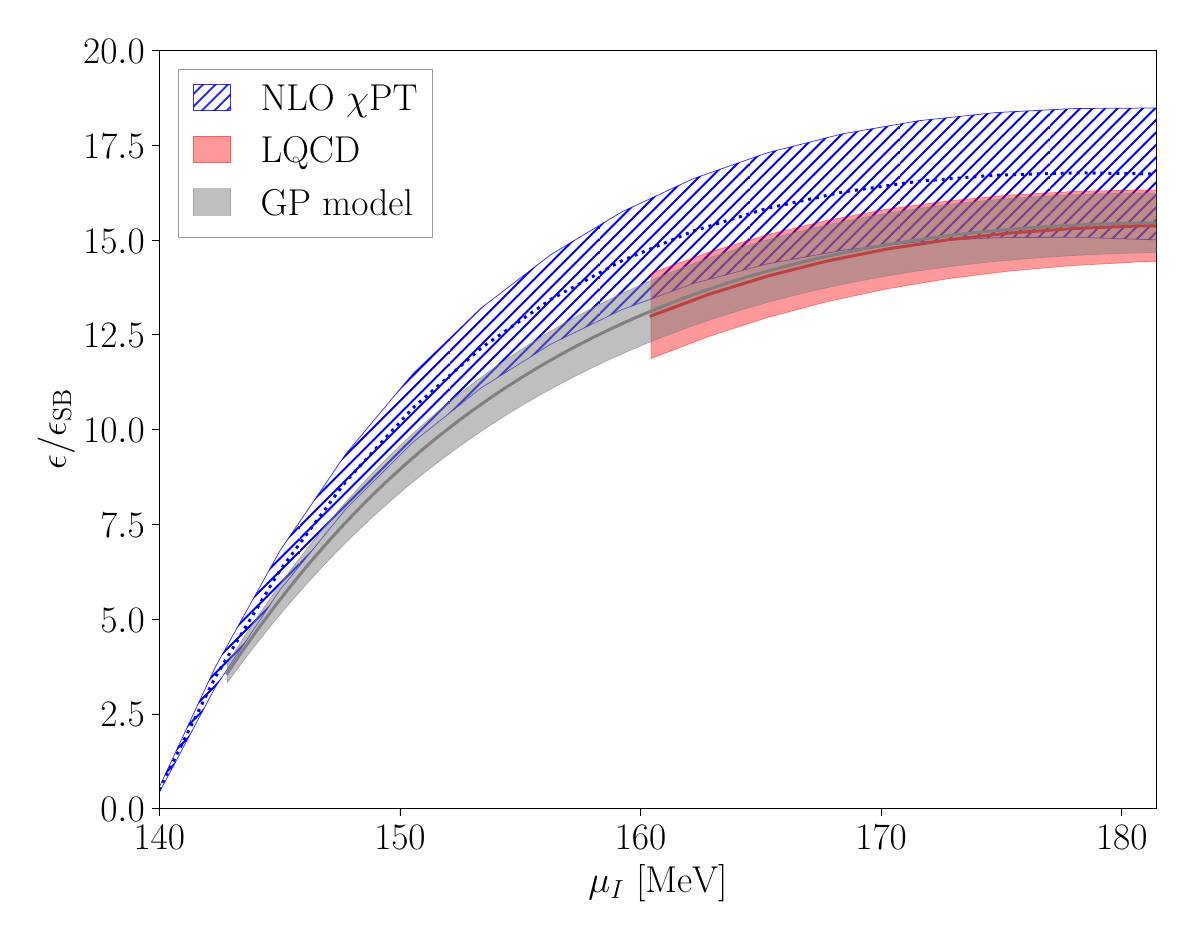}
    }
    \caption{ Low $\mu_I$ region for the squared speed of sound (a), pressure ratio (b), and energy density ratio (c). The GP model is shown in grey, while LQCD data is the red band and NLO $\chi$PT is the hatched blue band. \label{fig:modelzoomed} } 
\end{figure}

In order to further investigate the small $\mu_I$ region, all three thermodynamic quantities are displayed over the range $\mu_I\in \overline{m}_\pi\times[1.0,1.3]$ in Fig.~\ref{fig:modelzoomed} using a linear scale, showing the GP model, the LQCD data, and the NLO $\chi$PT result. Agreement is observed between $\chi$PT and LQCD for $\epsilon/\epsilon_{\rm SB}$ and $c_s^2/c^2$. 
However for the pressure ratio, some tension is observed for $\mu_I\alt \mu_I^{\rm min}$. As discussed above, for $\mu_I\sim m_\pi$, the LQCD calculations are dominated by systems of only a few pions which are far from the thermodynamic limit and omitted from the GP model. Consequently, $\chi$PT (which is most reliable for small $\mu_I$) primarily determines the GP model in this region.

\subsubsection{Hyperparameter dependence  }

The model mixing procedure depends on the choice of several hyperparameters. However, since LQCD data is available over a wide region, and there is at least one theory constraint for each value of $\mu_I$, the resulting GP model is mainly driven by the available information and shows mild hyperparameter dependence.

\begin{figure}[t!]
\centerline{\includegraphics[width=\linewidth]{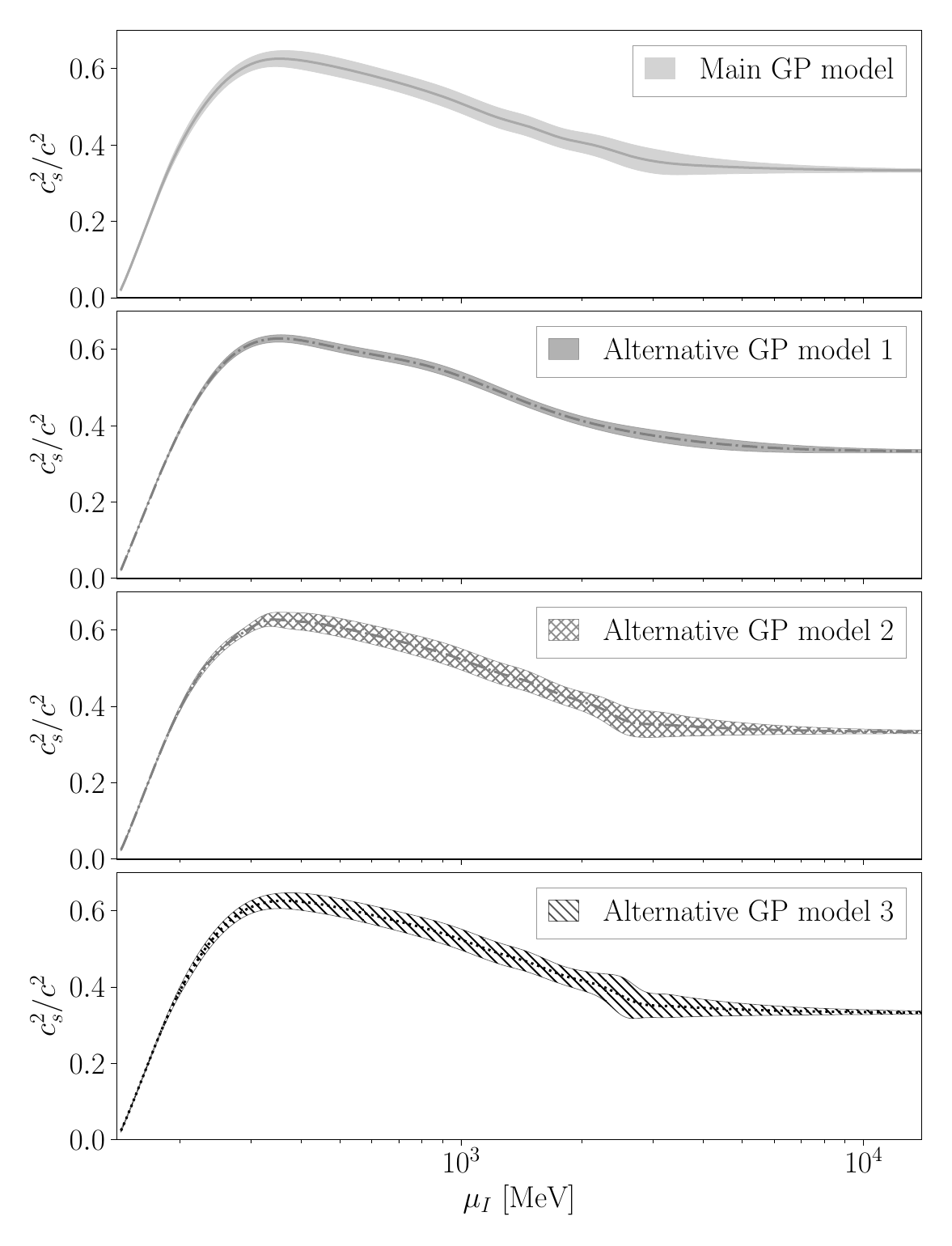}}
\caption{Squared speed of sound as a function of $\mu_I$ for the main GP model and three alternative choices for hyperparameters. 
}
\label{fig:speed-of-sound-several}
\end{figure}

In order to illustrate this, three additional alternative models are presented in Fig.~\ref{fig:speed-of-sound-several}:
\begin{itemize}
    \item {\bf Alternative Model 1}: Same as the main model, except that the prior, $\chi$PT and pQCD correlation lengths are chosen to be much larger, i.e., $\ell^{(nm)}_p=\ell^{(nm)}_{\chi {\rm PT}} = \ell^{(nm)}_{\rm pQCD} = 0.3$. This leads to a much more constrained GP model, but with consistent $\mu_I$ dependence. 
    
    \item  {\bf Alternative Model 2}: Same as the main model, except that the prior, $\chi$PT and pQCD correlation lengths are chosen to be much smaller, i.e., $\ell^{(nm)}_p=\ell^{(nm)}_{\chi {\rm PT}} = \ell^{(nm)}_{\rm pQCD} = 0.01$. This has almost no visible effect, except in the transition from LQCD to pQCD region, where the uncertainties become slightly larger.
    
    \item  {\bf Alternative Model 3}: Same as the main model, except that the range of validity of $\chi$PT is $\mu_I<1.5 \overline{m}_\pi$, and that of pQCD is $\mu_I>20 m_\pi$. This moderately increases  the uncertainties in the transition from LQCD to pQCD region.
\end{itemize}

\subsubsection{The effect of LQCD constraints}

\begin{figure}[t!]
\centerline{\includegraphics[width=\linewidth]{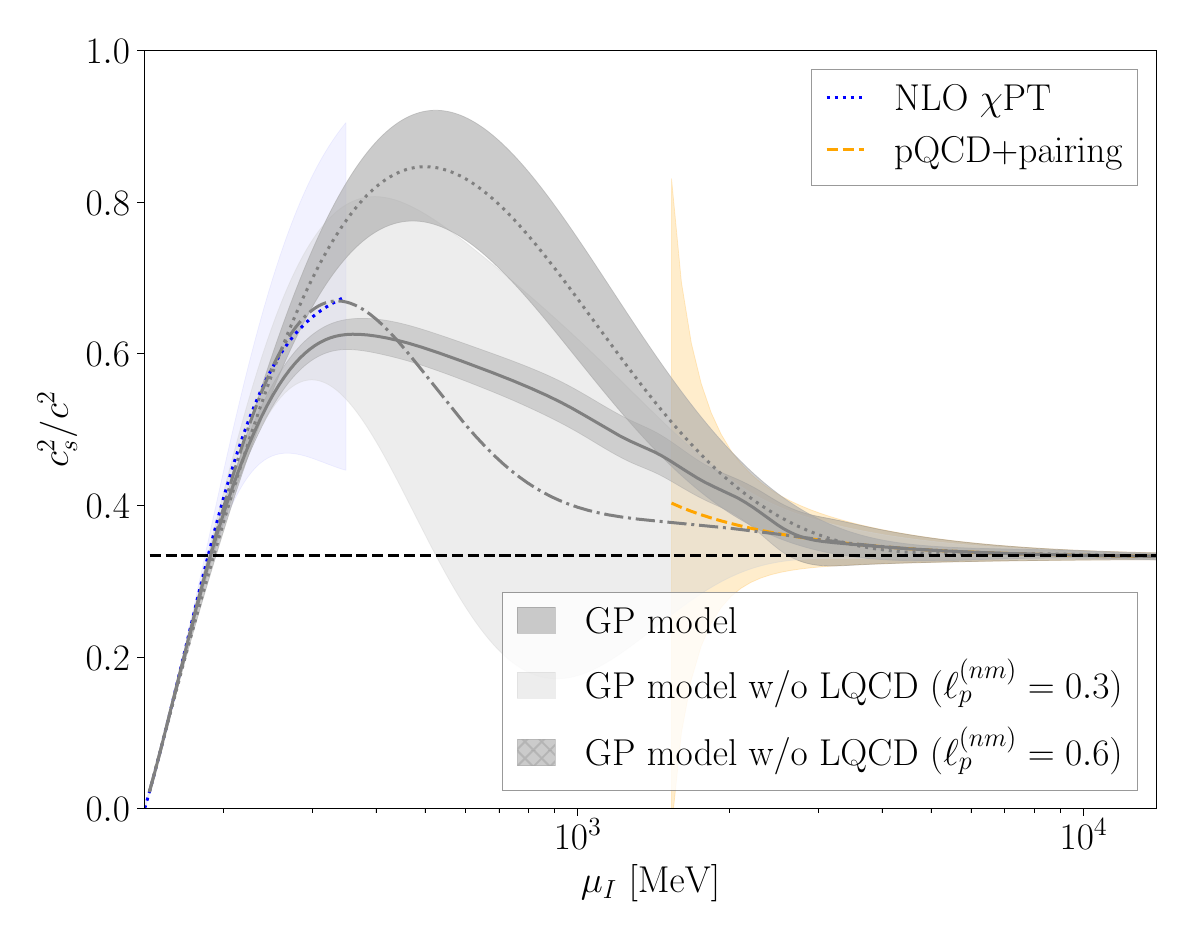}}
\caption{The squared speed of sound as a function of the isospin chemical potential. The darker grey solid line corresponds to the same GP model as in the main text. The lighter dot-dashed line corresponds to a GP model without LQCD inputs and a prior correlation length of $\ell^{(nm)}_p=0.3$. The dark dotted line is a GP model without LQCD inputs, but with a prior correlation length $\ell^{(nm)}_p=0.6$. The pQCD (orange), and $\chi$PT  (blue) constraints are also shown. The dashed horizontal line shows the conformal bound.
}
\label{fig:speed-of-sound-compare}
\end{figure}

In this section,  the importance of the LQCD results in constraining the EoS are discussed. In order to show this, three different GP-models are compared. 
One is the model in the main text with input from $\chi$PT, LQCD and pQCD, and using the hyperparameters described above. The other two GP-models do not contain LQCD constraints. In this case, due to the lack of information in the intermediate $\mu_I$ region, it is necessary to use a larger correlation length in the prior so that the posterior does not simply reproduce the input prior.
Specifically, two larger choices for the prior correlation length are used, $\ell^{(nm)}_p=0.3$ and $\ell^{(nm)}_p=0.6$, with the other hyperparameters being unchanged. The results for the speed of sound are shown in Fig.~\ref{fig:speed-of-sound-compare} for all three cases. 
As can be seen, the model including LQCD constraints and the model with $\ell^{(nm)}_p=0.3$ but no LQCD are consistent, however in the latter case, the uncertainties in the intermediate $\mu_I$ region are dominated by the choice of the prior.
In contrast, the GP model with $\ell^{(nm)}_p=0.6$ but no LQCD constraints produces a much more constrained interpolation in the intermediate region that is seen to be inconsistent  when compared with LQCD. The larger value of $\ell^{(nm)}_p$ implies stronger assumptions about the smoothness of the function, thus yielding more constrained posteriors, however those assumptions need not be consistent with the actual underlying function. Note that large correlation lengths are  used in similar GP-models for the EoS for non-zero baryon chemical potential, and the above variations suggest the EoS  in the unconstrained intermediate region may have strong model dependence.

\subsubsection{Example of GP model without gap}

\begin{figure*}[ht!]
     \centering
      \subfloat[]{
     \includegraphics[width=0.5\textwidth]{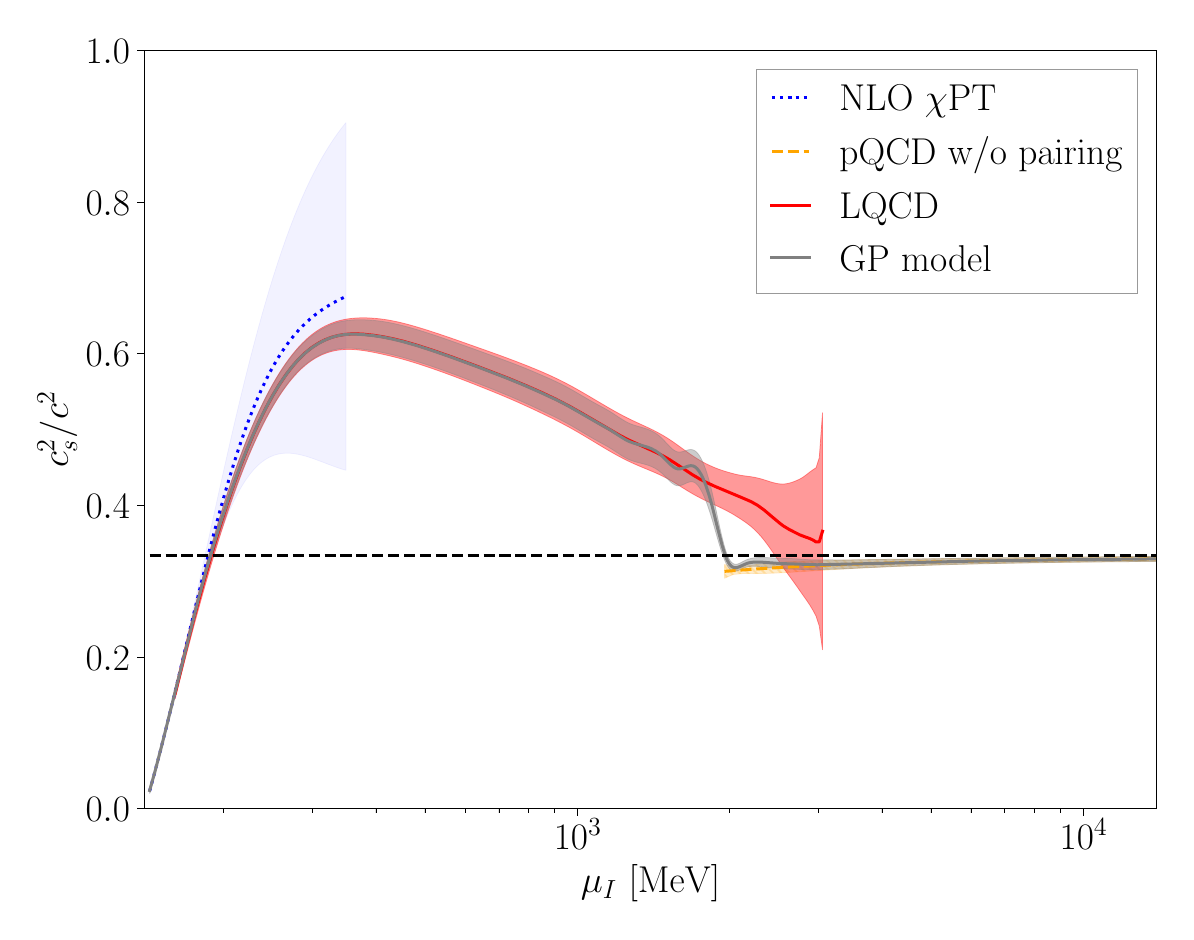}
    } 
     \subfloat[]{
     \includegraphics[width=0.5\textwidth]{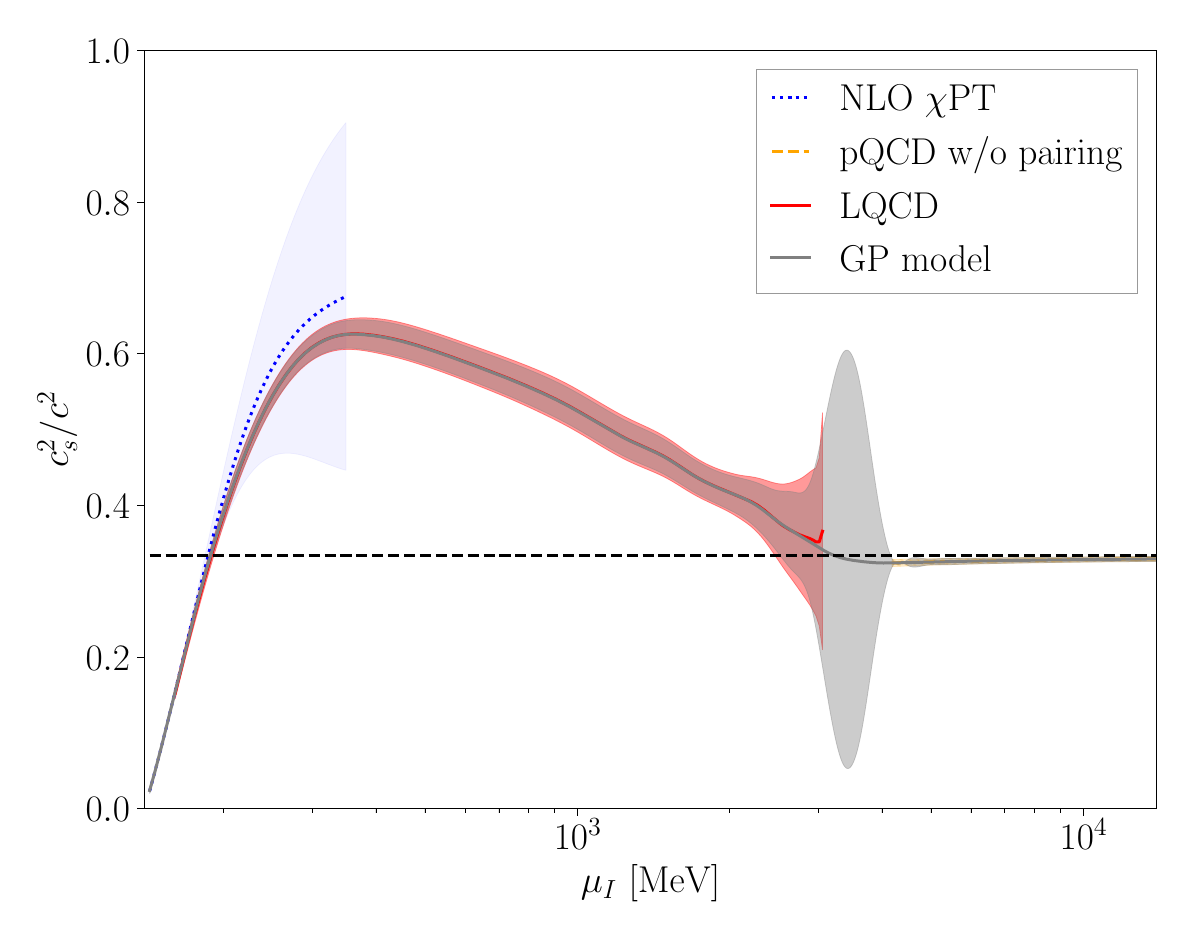}
    }
    \caption{Same as Fig.~\ref{fig:speed-of-sound}, but using NNLO pQCD without a pairing gap as a model constraint in the large $\mu_I$ region. The left panel (a) assumes that the range of validity of pQCD without a superconducting gap is $\mu_I>14 \overline{m}_\pi$, while the right panel (b) assumes this for $\mu_I>30 \overline{m}_\pi$.} 
    \label{fig:speed-of-sound-nogap}
\end{figure*}

So far, all the GP models have assumed that NNLO pQCD including the NLO pairing gap (Eq.~\eqref{eq:Delta-pert}) is the most accurate description at large $\mu_I$. 
However, an alternative explanation of the tension between the LQCD pressure and that of pQCD seen in Fig.~\ref{fig:pressure} is that the perturbative result only becomes valid at very large chemical potentials.
In order to test this, a GP model with the same hyperparameters as that of the main text, but with the pQCD with gap constraint replaced by pQCD without a gap is considered. The results are shown in Fig.~\ref{fig:speed-of-sound-nogap} for two choices of range of validity of pQCD. As can be seen, the transition between the lattice data and pQCD is more abrupt than when the pairing gap is considered, and significant tension between LQCD and pQCD is observed. This further supports the need to consider the pairing gap.

\subsection{Data file description}
In order to facilitate the use of the equation of state computed in this work, an HDF5~\cite{The_HDF_Group_Hierarchical_Data_Format} data file is provided, containing both the continuum-extrapolated LQCD data as well as the GP model derived from the LQCD data, $\chi$PT and pQCD.  At the top level, the data file is split into two groups, \mbox{\lstinline{/lattice}} and \mbox{\lstinline{/model_mixed},} with the former containing the continuum extrapolated lattice data, and the latter containing the model-mixed results.

For the continuum-extrapolated LQCD data, the \mbox{\lstinline{/lattice}} group contains bootstrap samples of the data, with the speed of sound squared stored in \mbox{\lstinline{/lattice/speed_of_sound2},} pressure stored in \mbox{\lstinline{/lattice/pressure},} and energy density stored in \mbox{\lstinline{/lattice/energy_density}.} Each of these quantities are arrays with dimensions $(N_\mu, N_\text{boot})$, where $N_\mu = 300$ is the number of isospin chemical potential values used, and $N_\text{boot} = 2000$ is the number of bootstrap samples. The isospin chemical potential values are given in \mbox{\lstinline{/lattice/chemical_potential}.} The isospin chemical potential is in units of $\mathrm{GeV}$, while the energy density and pressure are in units of $\mathrm{GeV}^4$ and the speed of sound is in units of $c$, the speed of light.

For the model-mixed data, the \mbox{\lstinline{/model_mixed}} group contains central values as well as a covariance matrix across the energy density, pressure, and speed of sound. The central values across all three quantities are stored in \mbox{\lstinline{/mixed_model/central_values},} while their covariances are in \mbox{\lstinline{/mixed_model/covariance}.} The values of the isospin chemical potential are given in \mbox{\lstinline{/mixed_model/logmuoverMpi},} which gives the values of $\log_{10} \mu_I^\text{phys} / \overline{m}_\pi$. The code sample below contains a snippet for obtaining samples of the speed of sound squared, pressure, and energy density in a NumPy array.

\vspace*{9cm}
\newpage

\begin{lstlisting}[language=Python,keepspaces=false]
import numpy as np
import h5py

with h5py.File("publish_data.h5", 'r') as data:
    central_values = np.array(data["mixed_model/central_values"])
    covariance = np.array(data["mixed_model/covariance"])
    evalpoints = np.array(data["mixed_model/logmuoverMpi"])

batch_size = 100
samples = np.random.multivariate_normal(
    central_values, covariance, size=batch_size)
samples = samples.reshape(batch_size, 3, 50)

mpi_phys_GeV = 139.57039 / 1000
mu_I = 10**evalpoints * mpi_phys_GeV

# Speed of sound squared
speed_of_sound_squared = samples[:,0]
# P / P_{SB}
pressure_ratio = samples[:,1]
# \epsilon / \epsilon_{SB}       
energy_ratio = samples[:,2]
\end{lstlisting}

\end{document}